\documentclass[sigconf]{acmart}
\AtBeginDocument{%
  \providecommand\BibTeX{{%
    \normalfont B\kern-0.5em{\scshape i\kern-0.25em b}\kern-0.8em\TeX}}}

\usepackage{etoolbox}
\newtoggle{vtwo}

\togglefalse{vtwo}

\newcommand{\secondversion}[2]{%
\iftoggle{vtwo}{%
{#1}
}
{#2}
}

\toggletrue{vtwo}     
\setcopyright{acmcopyright}
\copyrightyear{2022} 
\acmYear{2022} 
\setcopyright{rightsretained} 
\acmConference[MobiHoc '22]{The Twenty-third International Symposium
on Theory, Algorithmic Foundations, and Protocol Design for Mobile
Networks and Mobile Computing}{October 17--20, 2022}{Seoul, Republic
of Korea}
\acmBooktitle{The Twenty-third International Symposium on Theory,
Algorithmic Foundations, and Protocol Design for Mobile Networks and
Mobile Computing (MobiHoc '22), October 17--20, 2022, Seoul, Republic
of Korea}\acmDOI{10.1145/3492866.3549713}
\acmISBN{978-1-4503-9165-8/22/10}

\usepackage{amsmath,amsthm}
\usepackage{natbib}
\usepackage{graphicx}
\usepackage{mathtools}
\usepackage{color}
\usepackage[labelformat=simple]{subcaption}
\usepackage{pbox}
\usepackage{enumitem}
\usepackage{url}
\usepackage{dirtytalk}
\usepackage{bbm}
\usepackage{todonotes}
\usepackage{algpseudocode}
\usepackage{algorithm}
\usepackage{bookmark}
\usepackage{cleveref}
\usepackage{nicefrac}
\usepackage[normalem]{ulem}
\usepackage{bm}
\usepackage[export]{adjustbox}
\usepackage{float}
\usepackage{placeins}
\usepackage{thmtools}
\usepackage{thm-restate}
\usepackage{lipsum}% http://ctan.org/pkg/lipsum
\makeatother
\allowdisplaybreaks

\newtheorem*{remark*}{Remark}

\newcommand{\ALOOP}[1]{\ALC@it\algorithmicloop\ #1%
  \begin{ALC@loop}}
\newcommand{\ENDALOOP}{\end{ALC@loop}\ALC@it\algorithmicendloop}

\newcommand{\suppn}{{(n)}}
\newcommand{\recovery}{\bm{R}}

\newcommand{\servicecapacityset}{\Lambda}
\newcommand{\probabvector}{\bm{p}}
\newcommand{\Ex}[1]{\mathbb{E}\left[#1\right]}

\newcommand{\set}{\bm{S}}
\newcommand{\allocation}{\alpha}
\newcommand{\responsetime}{T^\suppn}

\newcommand{\psarrival}{\nu^\suppn}
\newcommand{\arrival}{\lambda^\suppn}

\newcommand{\numserver}{n^\suppn}

\newcommand{\numcoded}{n_{\text{coded}}^\suppn}
\newcommand{\bottleneck}{{i^*}}
\newcommand{\bottlenecktwo}{{k^*}}

\newcommand{\probab}{q^\suppn}
\newcommand{\slack}{\beta^\suppn}

\newtheorem{lem}{Lemma}
\newtheorem{thm}{Theorem}
\newtheorem{defn}{Definition}

\newtheorem{exple}{Example}

\let\oldexple\exple
\renewcommand{\exple}{\oldexple\normalfont}

\newtheorem{propty}{Property}

\graphicspath{{Figures/}}

\crefname{equation}{}{}
\Crefname{equation}{}{}
\crefname{thm}{theorem}{theorems}
\Crefname{thm}{Theorem}{Theorems}
\crefname{clm}{claim}{claims}
\Crefname{clm}{Claim}{Claims}
\Crefname{coro}{Corollary}{Corollaries}
\crefname{lem}{lemma}{lemmas}
\Crefname{lem}{Lemma}{Lemmas}
\crefname{coro}{corollary}{corollaries}
\Crefname{coro}{Corollary}{Corollaries}
\Crefname{sec}{Section}{Sections}
\crefname{app}{appendix}{appendices}
\Crefname{app}{Appendix}{Appendices}
\crefname{prop}{proposition}{propositions}
\Crefname{prop}{Proposition}{Propositions}
\Crefname{propty}{Property}{Properties}
\crefname{figure}{fig.}{figures}
\Crefname{figure}{Fig.}{Figures}
\crefname{defn}{definition}{definitions}
\Crefname{defn}{Definition}{Definitions}
\crefname{fact}{fact}{facts}
\Crefname{fact}{Fact}{Facts}
\crefname{appendix}{appendix}{appendices}
\Crefname{appendix}{Appendix}{Appendices}
\crefname{algo}{algorithm}{algorithms}
\Crefname{algo}{Algorithm}{Algorithms}
\crefname{algorithm}{algorithm}{algorithms}
\Crefname{algorithm}{Algorithm}{Algorithms}
\crefname{conj}{conjecture}{conjectures}
\Crefname{conj}{Conjecture}{Conjectures}
\crefname{obs}{observation}{observations}
\Crefname{obs}{Observation}{Observations}
\crefname{poli}{policy}{policies}
\Crefname{poli}{Policy}{Policies}
\crefname{exple}{example}{example}
\Crefname{exple}{Example}{Examples}

\begin{document}
\title{Tackling Heterogeneous Traffic in Multi-access Systems via Erasure Coded Servers}
\date{}
\author{Tuhinangshu Choudhury}
\affiliation{%
  \institution{Carnegie Mellon University}
  \city{Pittsburgh}
  \state{PA}
}

\author{Weina Wang}
\affiliation{%
  \institution{Carnegie Mellon University}
  \city{Pittsburgh}
  \state{PA}
}

\author{Gauri Joshi}
\affiliation{%
  \institution{Carnegie Mellon University}
  \city{Pittsburgh}
  \state{PA}
}

\begin{abstract}
Most data generated by modern applications is stored in the cloud, and there is an exponential growth in the volume of jobs to access these data and perform computations using them. The volume of data access or computing jobs can be heterogeneous across different job types and can unpredictably change over time. Cloud service providers cope with this demand heterogeneity and unpredictability by over-provisioning the number of servers hosting each job type. In this paper, we propose the addition of erasure-coded servers that can flexibly serve multiple job types without additional storage cost. We analyze the service capacity region and the response time of such erasure-coded systems and compare them with standard uncoded replication-based systems currently used in the cloud. We show that coding expands the service capacity region, thus enabling the system to handle variability in demand for different data types. Moreover, we characterize the response time of the coded system in various arrival rate regimes. This analysis reveals that adding even a small number of coded servers can significantly reduce the mean response time, with a drastic reduction in regimes where the demand is skewed across different job types.

\end{abstract}



\maketitle
\section{Introduction}
Modern-day cloud computing systems are used for various big-data applications such as performing  machine learning (ML) inference tasks \cite{dally2015high}, hosting large files for web services \cite{shvachko2010hadoop}, computing the PageRank of web graphs \cite{page1999pagerank}, and other data-intensive computations. Since these jobs require access to the specific data stored on the cloud server(s), each server is dedicated to one job type depending on the availability of data and high-performance hardware required for it. Therefore, the massive volume of users performing such cloud computing jobs have to contend for the server(s) storing data relevant to their job. For example, an ML inference system may store one trained model on each server, users that need to perform inference using that model are directed to that server. Or a cloud storage system hosting files for OTT platforms such as Netflix might use each server to store one movie, and users requesting that movie are assigned to that server. Therefore, the massive volume of users performing different cloud computing jobs have to contend for the server(s) storing data relevant to their job.

The traffic for different job types can vary due to diurnal variations or unpredictable demand fluctuations. These traffic variations often exhibit a negative correlation, i.e., if one job type is experiencing high traffic, another job type experiences low traffic. For e.g., in an ML inference system hosting different specialized models to process pictures captured by users, the inference traffic for each model can vary periodically depending on the time zone of the users accessing them. Such computing and inference jobs are latency-sensitive. Thus, a larger number of servers needs to be allocated to job types experiencing high traffic to satisfy the latency requirements of those users. The ideal solution for such situations is to enable dynamic allocation of servers where the number of servers provided to a job type depends on its current traffic. However, the servers often host large files (for e.g., the size of the Google Web crawler is over $10$ million GB) and have specialized computing hardware. Dynamically reconfiguring a server for a different job type would require the movement of large amounts of data and may not even be possible due to hardware constraints. Hence, there is a critical need to design multi-access computing systems resilient to traffic variations that avoid dynamic reconfiguration of servers.

One standard solution to handle traffic variations is to overprovision the number of servers dedicated to various job types to meet their peak demand. Overprovisioning can be achieved by adding replicas of servers dedicated to a job type \cite{CIRNE_comp_task_repl, sun2017delayoptimal} in proportion to the maximum historical demand for that job type over a large time horizon. While over-provisioning can meet latency requirements under traffic variations, it comes at the cost of severe underutilization of expensive computing resources and a massive energy footprint. Another solution is to supplement job-type-specific servers with flexible general servers that can serve more than one job type. However, adding such servers can be expensive because they need access to data, memory, and computing capabilities relevant to multiple job types. One such model was studied by \citeauthor{tsitsi_flexible_queueing} in \cite{tsitsi_flexible_queueing} which showed that the addition of a small number of flexible servers could give a dramatic reduction in the queueing delay.

While the solutions mentioned above provide some robustness to variations in traffic, none of them considers the correlation between traffic. For negatively correlated arrival rates, the overall traffic in the system varies slowly over time, even though individual jobs may experience significant changes. A novel approach that is well-suited for such skewed traffic patterns was proposed in \cite{coded_queueing_mehmet, coded_queueing_mehmet_journal}, where the authors supplement replicas of the servers of each type by a set of \emph{erasure coded servers}. When a job is sent to an erasure-coded server, the output is a linear combination of the outputs of two or more of the regular servers. For example, in a matrix computation task where job type $1$ (or type $2$) seeks to compute the product $\bm{A} \bm{x}$ (or $\bm{B} \bm{x}$) of an incoming vector $\bm{x}$ with the matrix $\bm{A}$ (or $\bm{B}$) stored at a server, a coded server can store $\bm{A} + \bm{B}$ such that its output is $(\bm{A}  +\bm{B}) \bm{x}$. Thus, a type $1$ job can be served and its output $\bm{A} \bm{x}$ can be obtained using an uncoded server storing $\bm{B}$ and a coded server storing $\bm{A} + \bm{B}$. Using this property, erasure-coded servers can serve multiple job types when combined with regular servers. Furthermore, unlike the flexible servers proposed in \cite{tsitsi_flexible_queueing}, the coded servers do not require extra resources such as memory. For various encoding schemes, \cite{coded_queueing_mehmet, coded_queueing_mehmet_journal} showed an improvement in the service capacity region of the system, the set of arrival rates for which the system is stable. The addition of coded servers significantly expands the service capacity region, especially in regions where the traffic for different job types is negatively correlated.

\subsection{Main Contributions}

While \cite{coded_queueing_mehmet, coded_queueing_mehmet_journal} proposed the idea of using coded servers to handle traffic variations and showed an expansion of the service capacity region, these works did not analyze the impact of coded servers on performance metrics such as mean response time or tail latency. Latency is important as often the request for a service has a deadline, in which case just ensuring stability of the system might not satisfy the user requirements. Since a coded server has to be used with one or more other servers, coding can increase the system load and result in a higher response time in some traffic regimes. However, this effect on the mean response time is not yet well-understood. 

In this paper, we build upon the model provided in \cite{coded_queueing_mehmet, coded_queueing_mehmet_journal} and generalize it to consider $\geq 2$ job types and an arbitrary allocation of servers to each job type and the coded servers. We compare the coded system with an uncoded system with the same total number of servers and corroborate the insight that the coded system significantly improves the system's stability by increasing the volume of the service capacity region. In addition, we characterize the mean response times of the coded and uncoded systems in several traffic regimes. We show that for a large number of servers, our coded system has a comparable or significantly smaller mean response time in most traffic regimes. The reduction in mean response time is prominent when the traffic of various job types is heavy and skewed, i.e., some job types are experiencing significantly higher traffic than other job types. The only regime where the uncoded system is better is when all job types simultaneously have high traffic, which is unlikely to occur in practical applications. Thus, we show our coded system can better handle heterogeneous traffic for different job types, and it improves the stability as well as the latency of multi-access systems.

\subsection{Related Work}\label{sec:existing_approach}

Improving the latency and service capacity of multi-access cloud systems has been extensively studied in the literature. In the context of storage systems, it has been studied in the caching literature \cite{ali_coding_for_caching, breslau_web_caching, rabinovich2002web}, where the number and location of replicas are dynamically adjusted based on the traffic. However, in data-intensive computing and content access jobs, such dynamic reconfiguration of servers is not feasible. Another approach proposed in \cite{tsitsi_flexible_queueing} is to add flexible servers that can serve multiple job types, but these servers will require additional memory and computing capabilities. Our proposed coding strategy does not require dynamic server reconfiguration or expensive flexible servers.

The use of erasure coding in distributed storage and computing systems is not new, in most prior works, the purpose of coding is straggler mitigation in jobs with parallel computing tasks, or latency reduction by launching redundant replicas of a job. For jobs that are divided into many parallel tasks such as MapReduce workloads, the tail latency of waiting for the slowest task(s) can be reduced by replication of tasks \cite{dean2013tail, ananthanarayanan2013effective, wang2019efficient, joshi2021synergy, yasaman_dynamic_heterogeneity_icnp_2018} such that the completion of only a subset of tasks is sufficient to complete the job. A generalized form of replication is erasure coding using $(n,k)$ maximum-distance-separable (MDS) codes, which offers a more efficient method to mitigate stragglers. In the context of content download from distributed storage, erasure coded systems divide a file into $k$ chunks that are coded into $n$ chunks such that downloading any $k$ out of $n$ is sufficient to recover the file. The latency of such distributed systems with redundant requests is analyzed in \cite{chen_queue_meets_code, shah2016when, gaurij_eff_red, li_mean_field_cod, kristen_mor_alan:redun-d, joshi2021synergy}. All the above works consider homogeneous jobs of only one type. In contrast, this paper employs erasure coding to achieve a different goal of handling skews in the traffic of heterogeneous jobs. To the best of our knowledge, this novel application of coding has only been previously studied in \cite{coded_queueing_mehmet, coded_queueing_mehmet_journal}, which we build on by considering a more general system and characterizing the mean response time in addition to the service capacity region. 

\section{Problem Formulation}\label{sec:prob_form}
We consider a multi-access cloud system consisting of $n$ servers that are used to provide service to $k$ types of jobs. 
Jobs of each type~$i$ arrive into the system according to a Poisson process. We assume that the traffic variation happens at a slower time-scale such that the system reaches a steady state before the traffic pattern changes. Therefore, it suffices to consider the setting where type $i$ jobs have a constant arrival rate $\lambda_i$. The service times of $n$ servers are unit-rate exponential random variables and are i.i.d. across servers and the jobs assigned to the server.

\paragraph{\textbf{Uncoded System.}} In most current implementations, the $n$ servers are divided into $k$ disjoint subsets $\set_i,i=1,2,\dots,k,$ each consisting of $|\set_i| = n_i = \alpha_i n$ servers for fractions $0 < \alpha_i < 1$ such that $\sum_{i=1}^{k} \alpha_i = 1$. The fraction $\alpha_i$ of servers dedicated to type $i$ jobs is a hyperparameter that can be set by the system designer. For example, $\alpha_i$ can be proportional to the long-term average of past values of the arrival rate $\lambda_i$. Since future arrival rates are not known beforehand or they may vary with time, the number of servers dedicated to each job type may not be sufficient to satisfy its arrival rate. Dynamically re-configuring a server to serve a different job type is expensive and slow in cloud systems, because it involves the movement of large amounts of data. Thus, we consider that the server types are fixed beforehand and cannot be modified on the fly, that is, type $i$ systematic servers cannot be quickly reconfigured to serve jobs of type $j$ for $j \neq i$.

\begin{figure}[t]
    \centering
    \includegraphics[page = 9, trim = {5mm 12mm 5mm 5mm}, clip,  width=0.99\linewidth]{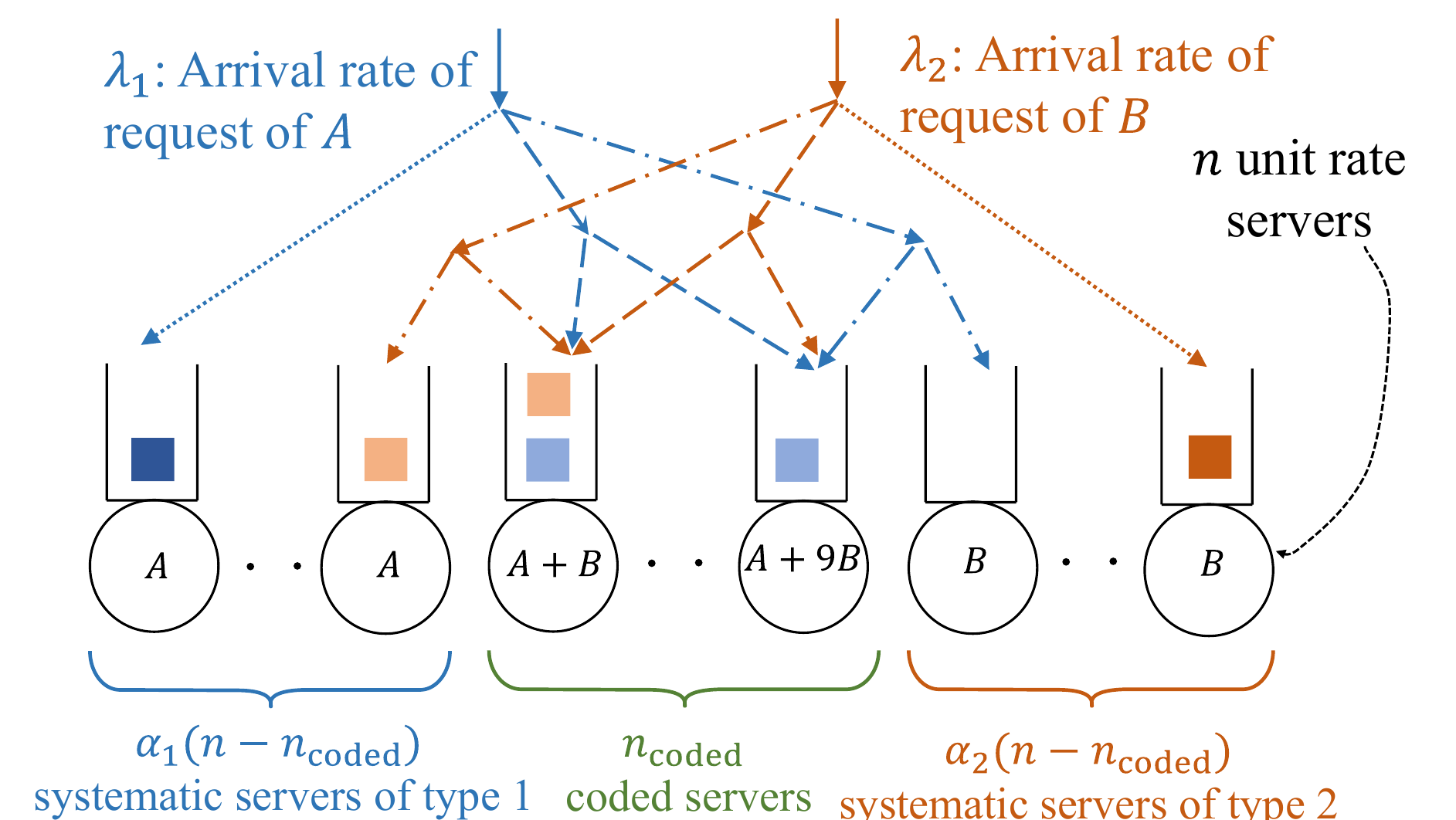}
    \label{fig:compare_privacy_heavy_arrival}
    \caption{
    An example of a multi-access computing system hosting two matrices $\bm{A}$ and $\bm{B}$, and consisting of $n$ unit-rate servers of which $n_{\text{coded}}$ are coded servers. The value written inside each server shows matrix stored in the server. Tasks corresponding to $A$ is represented by blue squares, while the one for $B$ is shown using orange squares. Different line types represents different ways in which a job can be served. 
    }
    \label{fig:sys_desc_cd}
\end{figure}

\paragraph{\textbf{Coded System.}}
In this paper, we consider a more generalized setup where out of $n$ servers, we reserve a set $\set_{\text{coded}}$ of  $n_{\text{coded}}\triangleq|\set_{\text{coded}}|$ servers to host \emph{erasure coded} versions of the job types. The remaining $n-n_{\text{coded}}$ systematic servers are split across the $k$ job-types such that $\alpha_i (n-n_{\text{coded}})$ servers are the systematic set $\set_i$ of job type $i$, with $0 < \alpha_i < 1$ for all $i = 1, \dots, k$ and $\sum_{i=1}^{k} \alpha_i =1$. We illustrate this server assignment for the $k=2$ case in \Cref{fig:sys_desc_cd}.

Erasure codes \cite{berkelamp1968algebraic} were originally developed in communication systems to add redundancy to transmitted messages in order to provide resilience against noise in the communication channel. In this paper, we employ them for a new purpose -- to handle skews in the arrival rate of various job types. When the arrival rate $\lambda_i$ of type $i$ jobs exceeds the cumulative service rate of the type $i$ systematic servers, the excess arrivals can be served using a combination of the coded servers and systematic servers of type(s) $j \neq i$. To illustrate the benefit of coded servers, we first give two concrete examples of erasure coded systems for $k=2$ job types. Then we describe the proposed coded system for general $k$.

\begin{exple}(Coded storage system)
\label{exple:coded_storage}

Consider an online storage platform with two video files, $V_1$ and $V_2$ of equal size, and $3$ servers that can host one file each. Consider that the servers store $V_1, V_2$ and $(V_1 \oplus V_2)$ respectively, where $(V_1 \oplus V_2)$ is the bit-wise XOR of $V_1$ and $V_2$. Note that the data-size stored on the coded server $(V_1 \oplus V_2)$ is the same as the file size $V_1$ and $V_2$. The jobs of type $1$ and $2$ are download requests for files $V_1$ and $V_2$. A type $1$ job can be served in two ways: 1) it can be sent to the systematic server storing $V_1$, or 2) it can be sent to the systematic server storing $V_2$ and the coded server storing $(V_1 \oplus V_2)$, and after downloading these two files, $V_1$ can be recovered by taking their bit-wise XOR, $(V_1 \oplus V_2)\oplus V_2 = V_1$. Thus, if the server storing $V_1$ cannot meet the demand for file $V_1$, the excess requests can be served using $V_2$ and $(V_1 \oplus V_2)$.
    
\end{exple}

To design more general coded storage systems with more than two files and more than one coded server, the files can be represented in a higher alphabet size than bits which will allow more flexible linear combinations. Such erasure coding of files is currently used in commercial cloud storage systems such as RAID \cite{chen1994raid} for the purpose of resilience against disk failures.

Besides coded storage systems where the jobs represent download requests, erasure coding can be applied to computing systems, as illustrated in \Cref{exple:coded_computing} below.

\begin{exple}(Coded computing system)
\label{exple:coded_computing}
Consider an online computing system, where a computing job seeks to find the product of an input vector $\bm{x}$ with the matrix $\bm{A}$ (or $\bm{B}$). Matrix computations are backbone of ML inference systems and such jobs are common where the vector $\bm{x}$ is an inference query and the matrix-vector product $\bm{A} \bm{x}$ (or $\bm{B} \bm{x}$) is the predicted output of a linear model. We assume that the matrices to be of same shapes. For different shapes, one can pad zeros to the matrices to unify their shapes.  Suppose we have a system consisting of $5$ servers, where each server stores one of two large matrices $\bm{A}$ and $\bm{B}$, or their linear combination. Consider that the $5$ servers store $\bm{A}, \bm{B}, (\bm{A} + \bm{B}), (\bm{A} - \bm{B})$, and $(\bm{A} + 2\bm{B})$, respectively. Given an input vector $\bm{x}$, the server multiplies the vector with the stored matrix and outputs the matrix-vector product. A type $1$ job that seeks to find the product $A \bm{x}$ can be served in the following ways: 1) sending the job to the server storing $\bm{A}$, 2) sending the job to any two of the three coded servers $(\bm{A} + \bm{B})$, $(\bm{A} - \bm{B})$ and $(\bm{A} + 2\bm{B})$ and obtain $\bm{A x}$ by taking a linear combination of the resulting matrix-vector products, or 3) sending the job to the systematic server storing $\bm{B}$ and any one of the three coded servers, and solving for $\bm{Ax}$ from the resulting matrix-vector products.

Since erasure codes are inherently linear, the coded computing framework described above can be directly applied only to linear computations such as matrix-vector multiplication. However, some recent research in coding theory is designing ways to apply erasure codes to non-linear computations \cite{ kosaian2019parity, mallick2021rateless} such as kernel methods and neural networks. The queueing and scheduling insights presented in this paper can be extended to the non-linear coded computing frameworks proposed in these works.
\end{exple}

\paragraph{\textbf{Maximum-Distance Separable Codes and Recovery Sets.}} The examples shown above considered just $2$ types of jobs. More generally, when there are $k$ types of jobs, we propose a coded multi-access system that employs a class of erasure codes called \emph{maximum-distance-separable (MDS) codes} \cite{berkelamp1968algebraic}. MDS codes are often used in distributed storage systems to provide resilience against disk failures. If there are $k$ files that need to be stored on $l$ disks, an $(l,k)$ MDS code constructs $l$ independent linear combinations of the $k$ files such that a file $i$ can be recovered from any set of $k$ coded files. A commonly used MDS code is the Reed-Solomon code, which constructs the linear combinations by evaluating a $k-1$-th degree polynomial at $l$ points. A special case of an $(l,k)$ MDS code is the \emph{systematic} MDS code, where $k$ of the $l$ combinations are uncoded copies of the $k$ files, and the remaining $l-k$ are other independent linear combinations. In our coded multi-access cloud system, we consider such a systematic MDS code. We have $|\set_i| = \alpha_i (n - n_\text{coded})$ servers storing uncoded copies of each job type $i$ and $n_{\text{coded}}$ independent linear combinations of the $k$ job types. 

In our MDS coded system, we can serve a type $i$ job using one of the following options: 1) one of the systematic servers from set $\set_i$, 2) any $k$ coded combinations from the set $\set_{\text{coded}}$ of coded servers, or 3) $k_1$ coded servers and $k-k_1$ distinct systematic servers from the sets $\set_j$, where $j\in\{1, 2, \cdots, k\}$\textbackslash $\{i\}$ and for some integer $k_1$ such that $1 \leq k_1 < k$. We denote the union of all these possible subsets of servers that can serve a type $i$ job by $\recovery_i$, and refer to it as the \emph{set of recovery sets}. The size $|\recovery_i|$ is the number of possible recovery sets for job type $i$, and $\recovery_{ij}$ denotes the set of servers corresponding to the $j$-th recovery set, for $j = 1, \dots, |\recovery_i|$. For instance, in \Cref{exple:coded_storage}, the set of recovery sets for the file $V_1$ is $\recovery_1 = \{ \{1 \}, \{2, 3\} \}$, consisting of the two possible combinations that can be used to recover $V_1$. 
\paragraph{\textbf{Queueing model.}} 
Next, we describe the queueing model for our system, which is also illustrated in \Cref{fig:sys_desc_cd}. We consider a first-come-first-served (FCFS) queue at each server. When a type $i$ job arrives, it needs to be assigned to a recovery set in $\recovery_i$ immediately. Specifically, if the chosen set consists of a single systematic server of type $i$, then the job needs to receive service from that systematic server, and thus we say that \emph{the job consists of one task}. If the chosen set contains $(k - k_1)$ systematic servers and $k_1$ coded servers for some $k_1 \geq 0$, then the job needs to receive service from all the $k$ servers, and thus we say that \emph{the job consists of $k$ tasks}. This service model induces the following queueing dynamics: when a job arrives, we send a task to the queue at each server in the chosen set in $\recovery_i$. The job is completed when all of its tasks are completed. We call the policy that determines which recovery set to assign to each arriving job as the \emph{routing policy}.

\paragraph{\textbf{Performance Metrics}} Since the coded system provides the flexibility of having multiple ways of serving a job, it improves load balancing. However, this flexibility comes at the cost of redundancy because to serve a job using a coded combination, we need responses from $k$ servers. To compare the coded and uncoded systems in terms of flexibility vs redundancy trade-off, we use two performance metrics: $(1)$ the service capacity region, the set of arrival rate vectors for which the system is stable, and $(2)$ the mean response time of jobs. We define these metrics formally below.

\begin{defn}[Service Capacity Region]\label{def:capacity_region}
The service capacity region of a multi-access cloud system, denoted by $\servicecapacityset$, is the region such that for any arrival rate vector $\bm{\lambda} = (\lambda_1, \cdots,  \lambda_k)$ in the interior of the region, there exists a routing policy under which 
\begin{equation}
    \lim_{c\rightarrow \infty}\lim_{t\rightarrow \infty}\mathbb{P}\left(\text{Number of jobs in the system at time $t$}> c\right) = 0.
\end{equation}
\end{defn}
The service capacity region consists of all supportable throughput.  Therefore, it is a notion independent of policies, and it measures the fundamental limit of the system.

\begin{defn}[Mean Response time]\label{def:mean_resp_time} 
The \emph{response time} $T$ of a job is the total time that a job spends in the system from its arrival until all its tasks are completed. If the system is stable, then the \emph{mean response time} of the system, denoted by $\Ex{T}$, is defined as follows when the limit on the right-hand-side exists with probability $1$:
\begin{equation}
    \Ex{T} = \lim_{m\rightarrow \infty}\frac{1}{m}\sum_{j = 1}^m T_j.
\end{equation}
where $T_j$ denotes the response time of the $j$-th job that departs from the system. For an unstable system, we define the mean response time $\Ex{T}$ to be $\infty$.
\end{defn}
The mean response time is a performance metric specific to the routing policy. In this paper, we consider probabilistic job routing policies. 
In particular, for any $\recovery_i$, we fix a probability vector $\probabvector_i$ of length $|\recovery_i|$, and any incoming job of type $i$ is assigned to $\recovery_{ij}$ with probability $p_{ij}$, the $j$-th element of $\probabvector_i$. Then, the total arrival rate of type-$i$ job to the set $\recovery_{ij}$ is given by $\lambda_{ij} = \lambda_ip_{ij}$. 
When comparing the mean response times of the coded and uncoded systems, we compare the best achievable performances by considering the optimal probabilistic routing policy that minimizes the mean response time. For example, for the uncoded system, the optimal routing policy corresponds to sending a job of type $i$ to one of the systematic servers of type $i$ chosen uniformly at random. Obtaining the optimal routing policy for the coded system can be difficult. We provide some routing policies that perform well, both theoretically and practically in \Cref{sec:latency_charac} and \Cref{sec:simulations}.

For the analysis of the mean response time, we assume that the number of coded servers is  $n_{\text{coded}} = o(n)$. The case of $n_{\text{coded}} = \Theta(n)$ is both feasible and interesting. However, the comparison of the mean response times of the uncoded and coded systems becomes intractable in this regime. That is why we focus on $n_{\text{coded}} = o(n)$ for the response time characterization presented in \Cref{sec:latency_charac}. However, we provide some insights for $n_{\text{coded}} = \Theta(n)$ regime in our simulations in \Cref{sec:simulations}. 

We also assume that $\lambda_i = \Theta(n)$ for all $i$. We believe that this is a mild assumption because the traffic experienced by a job is usually proportional to the servers allocated to the job type.

\paragraph{\textbf{Organization of the paper.}} 
In \Cref{sec:service_capacity_region}, we analyze the service capacity region of the coded and uncoded system. \Cref{sec:latency_charac} compare the mean response times of the coded and uncoded system. In \Cref{sec:simulations_far}, we present simulation results that corroborate our theoretical result given in \Cref{thm:rt_analyze_general} for a fixed arrival rate vector. In \Cref{sec:simulations_var}, we provide simulations with variable arrival rate vectors and show how negative correlation in traffic benefits the coded system. Finally, in \Cref{sec:conclusion} we provide a conclusion and directions for future work.

\section{Service Capacity Region}\label{sec:service_capacity_region}
The service capacity region is the set of job arrival rates for which the system is stable, that is, the cumulative arrival rate to any server does not exceed its service rate, as defined in \Cref{def:capacity_region}.  \Cref{lem:capacity_region_uncoded} and \Cref{lem:capacity_region_coded} below provide the service capacity regions for the uncoded and coded multi-access cloud systems. We show that the service capacity region of the coded system is significantly larger than that of the uncoded system.

\begin{figure}[t]
\begin{subfigure}{0.52\linewidth}
    \centering
    \includegraphics[page = 8, trim = {7mm 4mm 9mm 9mm}, clip,  width=1\linewidth]{newFigures/newFigure.pdf}
    \caption{$k=2$.}
    \label{fig:service_cap_2k}
\end{subfigure}
\hfill
\begin{subfigure}{0.47\linewidth}
    \centering
    \includegraphics[trim = {45mm 80mm 45mm 80mm}, clip,  height = .9\linewidth, width=0.9\linewidth]{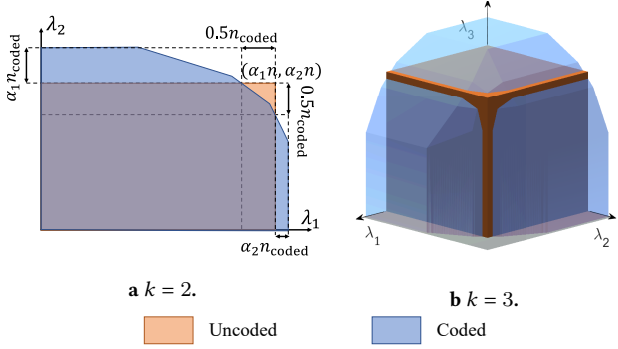}
    \label{fig:service_cap_3k}
    \caption{$k=3$.}
\end{subfigure}
\hfill
\centering
\begin{subfigure}{\linewidth}
    \centering
    \includegraphics[page = 4, trim = {6mm 13mm 106mm 10mm}, clip, width=0.54\linewidth]{newFigures/newFigure.pdf}
\end{subfigure}
\caption{ 
Service Capacity region of our system for $k=2, 3$. The blue region represents the service capacity region for the coded system, and the orange one shows the service capacity region for the uncoded system. The service capacity region of the coded system expands in regimes where the traffic is skewed. However, it loses some area when all job types have similar and high arrival rates.
}
\label{fig:service_cap}
\end{figure}

\begin{lem}[Uncoded Service Capacity Region]\label{lem:capacity_region_uncoded}
Consider an uncoded multi-access cloud system consisting of $n$ servers and $k$ job types, with $|\set_i| = \allocation_i n$ servers allocated to job type $i$. Then, the service capacity region is the set 
\begin{equation}\label{eq:servicecap_uncoded}
    \servicecapacityset_\text{uncoded} = \left\{(\lambda_1, \dots,  \lambda_k): 0\leq \lambda_i \leq \allocation_i n, \forall i\in\{1, \dots, k\}\right\}.
\end{equation}
\end{lem}
The service capacity region of the uncoded system is the hyper-cuboid of size $k$ and length of the $i$-th side being $ \allocation_i n$. This comes from the fact that the total service capacity of systematic servers of type $i$ is $ \allocation_i n$. The orange region in \Cref{fig:service_cap} illustrates the service capacity region of the uncoded system for $k=2, 3$.
Next, we characterize the service capacity region of the coded system in \Cref{lem:capacity_region_coded}. We first characterize the service capacity region of the coded system using \eqref{eq:servicecap_coded}. However, for any vector $\bm{\lambda}$, it is difficult to verify whether $\bm{\lambda}$ satisfies \eqref{eq:servicecap_coded} or not. Hence, we provide a simpler characterization via \eqref{eq:stability_condition_coded} that is a necessary and sufficient condition for stability.

\begin{thm}[Coded Service Capacity Region]\label{lem:capacity_region_coded}
Consider a coded multi-access cloud system consisting of $n$ servers and $k$ job types. Let $\recovery_{ij}$ be the $j$-th element of $\recovery_{i}$, the set of recovery sets of type $i$. Then, the service capacity region is the set
\begin{align}
    \servicecapacityset_{\text{coded}} &= \Big\{(\lambda_1, \dots,  \lambda_k): \exists \lambda_{ij} \geq 0, \mbox{ s.t. } \lambda_i = \sum_{j =1}^{ \mid\recovery_i \mid} \lambda_{ij}, \forall i\in\{1, \dots, k\},\nonumber\\
    &\qquad \sum_{i = 1}^k\sum_{j: \ell\in \recovery_{ij}} \lambda_{ij} \leq 1, \forall \ell\in \{1, \dots, n\}\Big\}\label{eq:servicecap_coded}
\end{align}
Let $r_i$ denote the \emph{residual capacity} for type $i$ jobs, given as $r_i \triangleq \allocation_i(n - n_{\text{coded}}) - \lambda_i$. Also define $r_{i}^{+} \triangleq \max\{r_i, 0\}$, and $r_{i}^{-} \triangleq -\min\{r_i, 0\}.$ Without loss of generality, assume that $r_1 \leq r_2\leq \dots  \leq r_k$. 
Then, an arrival rate vector $(\lambda_1, \dots,  \lambda_k)\in \servicecapacityset_\text{coded}$ if and only if
\begin{equation}\label{eq:stability_condition_coded}
    \min_{k_0 \in \{1, \dots ,k\}}\left\{\frac{n_{\text{coded}} + \sum_{i = 1}^{k_0} r_{i}^+}{k_0}\right\} \geq \sum_{i = 1}^k r_{i}^{-}.
\end{equation}
\end{thm}

The proof sketch of \Cref{lem:capacity_region_coded} is given in \Cref{sec:prf_sketch_capacity} below, and the complete proof is given in \secondversion{\Cref{prf:capacity_region}.}{\cite{choudhury2021_coded_queueing}.}

For the special case of $k=2$ and $3$, the blue regions in \Cref{fig:service_cap} illustrate the service capacity region of the coded system. The service capacity region for $k=2$ was previously derived in \cite{coded_queueing_mehmet, coded_queueing_mehmet_journal}\secondversion{ and is also included in \Cref{prf:capacity_region}.}{.} However, \cite{coded_queueing_mehmet, coded_queueing_mehmet_journal} did not consider the general case when the number of job types $k > 2$. 
From \Cref{fig:service_cap_2k}, observe that there is an expansion in the service capacity region in the top-left and bottom-right areas, the regions where the traffic is heavy and skewed. The routing policy reduces the traffic in the higher loaded systematic servers by redirecting some of their load to the coded servers and less loaded systematic servers. The induced load balancing then allows the traffic of a job type to extend beyond the capacity of its systematic servers, leading to an expansion in the service capacity region. 

However, the coded system loses portion of the service capacity region in the top right corner, where the traffic is heavy but not skewed. In this region, the traffic of all job types is usually greater than the total capacity of their systematic servers, hence the systematic servers are not enough on their own to serve all job types. In addition, the coded servers add a significant amount of redundancy which further increases the traffic, effectively making it unstable.

For $k=2$, the total area of the regions gained is $\Theta(nn_{\text{coded}})$ since the maximum possible arrival rate of a job of type $i$ can increase by $\Theta(n_{\text{coded}})$. Likewise, the area of lost region is $\Theta((n_{\text{coded}})^2)$ which makes the total area of service capacity region in the coded system to be $\left(\allocation_1\allocation_2 n^2 + \Theta(nn_{\text{coded}}) - \Theta((n_{\text{coded}})^2)\right)$. If the number of coded servers is small, i.e., $n_{\text{coded}} = o(n)$, then there is an effective gain in the service capacity region for the coded system. 

\subsection{Proof Sketch of Theorem \ref{lem:capacity_region_coded}}\label{sec:prf_sketch_capacity}
The proof of \Cref{lem:capacity_region_coded} contains two parts, the proofs of \eqref{eq:servicecap_coded} and \eqref{eq:stability_condition_coded}, and both follow a similar pattern. We first prove that if any arrival rate vector satisfies the property (the requirement in \eqref{eq:servicecap_coded} or \eqref{eq:stability_condition_coded}), there is a way to stabilize the system. We then prove that if the property is not satisfied, no policy can stabilize the system. Below, we present the proof sketches of \eqref{eq:servicecap_coded} and \eqref{eq:stability_condition_coded}. \secondversion{The detailed proofs are provided in \Cref{prf:capacity_region}.}{}

\paragraph{\textbf{Proof of equation \eqref{eq:servicecap_coded}}}
We first prove that if any arrival rate vector $\bm{\lambda}$ satisfies \eqref{eq:servicecap_coded}, then there exists a way to stabilize the system. By the definition of the set $\servicecapacityset_{\text{coded}}$ given in \Cref{lem:capacity_region_coded}, for any arrival rate vector $\bm{\lambda}\in\servicecapacityset_{\text{coded}}$, there exists a set of $\lambda_{ij}$'s that satisfies \eqref{eq:servicecap_coded}. One routing policy for this arrival rate vector is to serve a job of type-$i$ using servers in the recovery set $\recovery_{ij}$ with probability $(\lambda_{ij}/\lambda_i)$. Under this policy, the total arrival rate to $\ell$-th server is $\sum_{i = 1}^k\sum_{j: \ell\in \recovery_{ij}} \lambda_{ij}$, where the inner sum represents the total arrival rate of tasks corresponding to the job of type $i$. Equation \eqref{eq:servicecap_coded} ensures that for all $\ell\in \{1, \dots, n\}$, the total arrival rate to $\ell$-th server is less than $1$, the service rate of the $\ell$-th server, which implies that the system is stable. 

The other direction of the result, i.e., any $\bm{\lambda}$ outside $\servicecapacityset_{\text{coded}}$ makes the system unstable, can be proven using standard techniques involving the \emph{Strict separation theorem} and the \emph{Strong law of large numbers} (see  Chapter $4.2$ of \cite{srikant2014communication} for an example). \secondversion{We omit the proof here for conciseness.}{}

\begin{figure}[t]
    \centering
    \includegraphics[page = 6, trim = {14mm 16mm 6mm 12mm}, clip,  width=.99\linewidth]{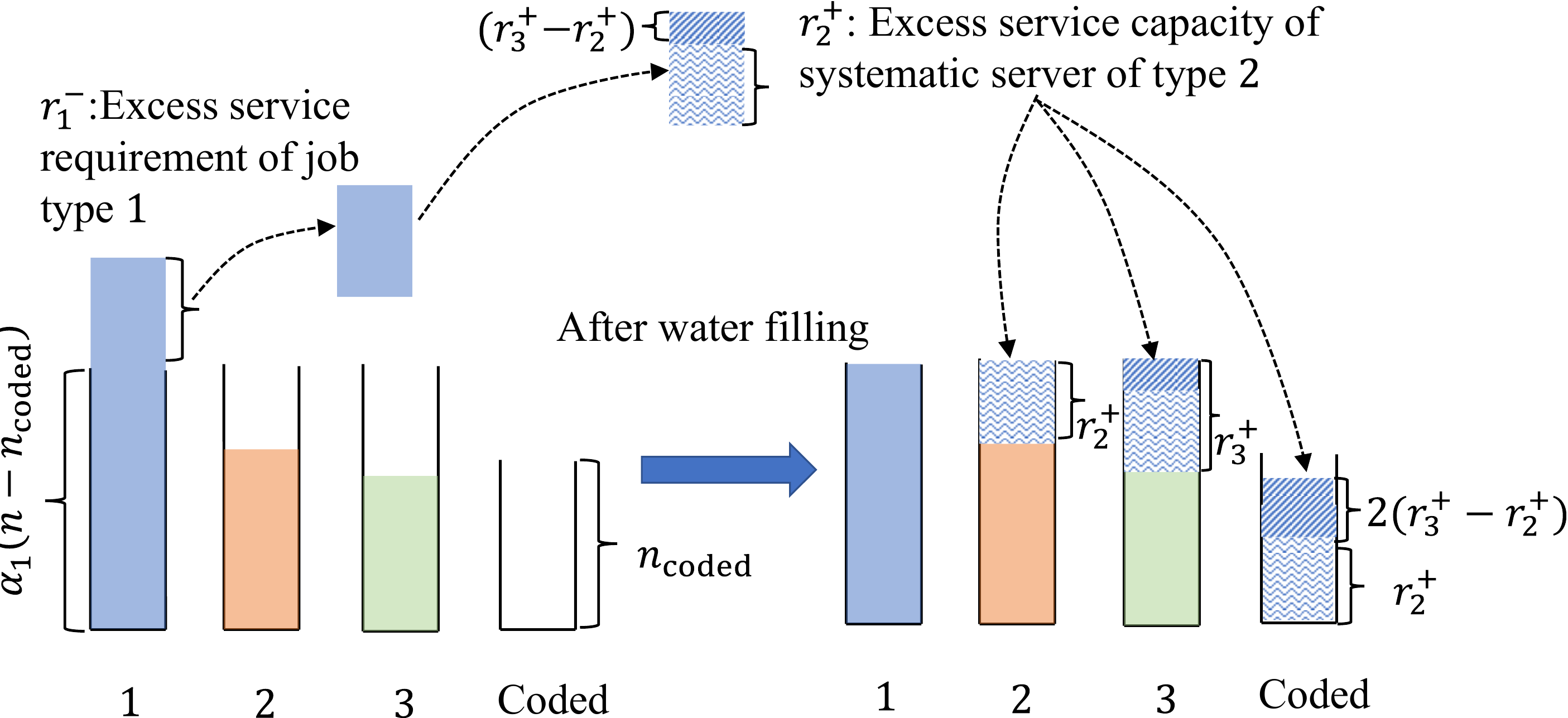}
    \caption{An illustration of the water filling method for $k = 3$ used in the proof of \eqref{eq:stability_condition_coded}. The water-filling strategy involves first filling the systematic servers to the brink. The remaining service requirement is then served using the coded servers and the under-utilized systematic servers.}
    \label{fig:waterfilling}
\end{figure}

\paragraph{\textbf{Proof sketch of equation \eqref{eq:stability_condition_coded}}}
We first prove that for any arrival rate vector satisfying \eqref{eq:stability_condition_coded}, there exists a policy that stabilizes the system. We use a water-filling argument where we initially serve a job only using its systematic servers. We start using the coded servers when the systematic servers reach their maximum capacity. Then, the right-hand side of \eqref{eq:stability_condition_coded} is the total excess service requirement after systematic servers are fully utilized. The left-hand side is the total service capacity available at the coded servers plus the excess service capacity at the underutilized systematic servers. The system is stable if the excess service capacity exceeds or equals the excess service requirement. A pictorial representation of the water-filling argument is given in \Cref{fig:waterfilling}.

We then show that every arrival rate vector $\bm{\lambda}\in\servicecapacityset_{\text{coded}}$ satisfies \eqref{eq:stability_condition_coded}. The key idea is that for any $\bm{\lambda}\in \servicecapacityset_{\text{coded}}$, there exist $\lambda_{ij}$'s satisfying \eqref{eq:servicecap_coded} such that for any $i$ either a job of type $i$ only use its own systematic servers or type $i$ systematic servers is used to serve only jobs of type $i$. This decomposition of $\bm{\lambda}$ ensures that the service requirement from the coded servers exceeds ($\sum_{i = 1}^k r_{i}^{-}$). Moreover, the service capacity available at the coded servers plus the under-utilized systematic servers is less than  $(n_{\text{coded}} + \sum_{i = 1}^{k_0} r_{i}^+)/k_0$ for any $k_0\in \{1, \dots, k\}$. The proof concludes by using the fact that the system is stable which implies equation \eqref{eq:stability_condition_coded} is true.

\section{Response time characterization} \label{sec:latency_charac}
\begin{figure}[t]
    \centering
    \includegraphics[page = 7, trim = {9mm 8mm 12mm 10mm}, clip,  width=\linewidth]{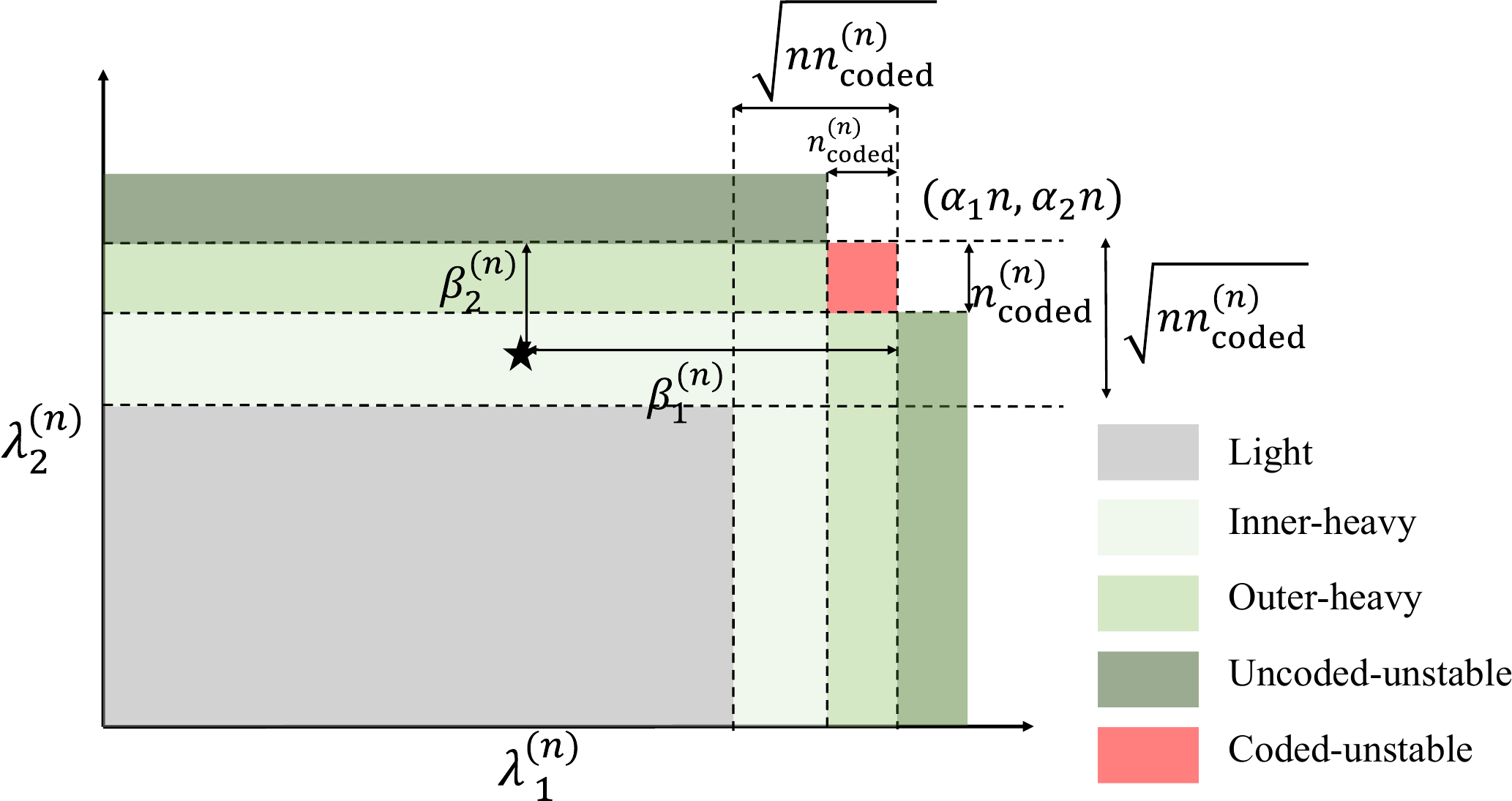}
    \caption{A pictorial representation of the traffic regimes for the case $k=2$. The star represents a possible value of arrival rate vector and $\slack_i$ is represented by it's distance from the edge. The response time comparison of the coded and uncoded system for these regimes is provided in \Cref{thm:rt_analyze_general}.
    }
    \label{fig:response_time_2_file}
\end{figure}

In this section, we compare the minimum mean response times of the coded and uncoded systems, denoted by $\Ex{\responsetime_\text{coded}}$ and $\Ex{\responsetime_\text{uncoded}}$ respectively. For the remainder of the paper, we use the notation superscript $(n)$ to denote that the quantity depends on $n$. To characterize and compare the response times, we consider five traffic regimes listed and illustrated in \Cref{fig:response_time_2_file}. We formally define these regimes in \Cref{sec:regions}, and then we compare the response times in each of these regimes in our main theorem in \Cref{sec:main_result}.

Note that \Cref{fig:response_time_2_file} illustrates the \emph{orders} of the edges of different areas as $n$ becomes large. Hence, the slanted edges in \Cref{fig:service_cap_2k} translates to vertical and horizontal lines in \Cref{fig:response_time_2_file}.

% --------------------------------------------------------------------------------------
\subsection{Traffic Regimes}\label{sec:regions}

Recall that jobs of type $i$ arrive at the system according to a Poisson process with rate $\arrival_i$. Let $\slack_i$, referred to as the \emph{slack capacity} of systematic servers of type $i$, be defined as
\begin{equation}
 \slack_i = \allocation_i n -\arrival_i.
\end{equation}
We define five traffic regimes based on the orders of the slack capacities of all job types. For ease of exposition, we first describe the regimes for the case where the system has two job types, i.e., $k = 2$, and then we generalize them to system with any value of $k$.

\subsubsection*{\textbf{Traffic regimes for $k= 2$}}\label{sec:regions_2}
Without loss of generality, we assume that $\slack_1 \leq \slack_2$.

\paragraph{Light regime:} $\slack_i\ge 0$ for all $i\in\{1,2\}$, and \[\slack_i = \omega\left(\sqrt{n\numcoded}\right),\quad \text{for all } i\in\{1,2\}.\]
In this regime, the slack capacities are large for both job types.

\paragraph{Inner-heavy regime:} $\slack_i\ge 0$ for all $i\in\{1,2\}$, and 
\begin{gather*}
\slack_1 = o\left(\sqrt{n\numcoded}\right), \slack_1 = \Omega\left(\numcoded\right), \slack_2 = \omega\left(\slack_1\right).
\end{gather*}

In this regime, type $1$ jobs experience heavier traffic than type $2$ jobs do, and thus type $1$ jobs can benefit from a coded system.

\paragraph{Outer-heavy regime:} $\slack_i \geq 0$ for all $i\in\{1,2\}$, and
\begin{gather*}
    \slack_1 = o\left(\numcoded\right),\slack_2 = \omega\left(\numcoded\right).
\end{gather*}
Compared to the inner-heavy regime, the outer-heavy regime has an even heavier traffic for type~$1$ jobs, so the traffic is further skewed.

\paragraph{Uncoded-unstable:} In this regime, the uncoded system is unstable while the coded system is stable.
\paragraph{Coded-unstable:} In this regime, the coded system is unstable while the uncoded system is stable.

\subsubsection*{\textbf{Traffic regimes for a general $k$}}\label{sec:regions_k}
Without loss of generality, we assume that $\slack_1 \leq  \dots \slack_k$. To generalize the definitions of the five traffic regimes, we divide the $k$ job types into two groups based on their slack capacities, $(1)$ \emph{beneficiaries}, and  $(2)$ \emph{helpers}.

Intuitively speaking, the beneficiaries are the job types that experience heavier traffic, and the helpers are those who experience lighter traffic.  Therefore, the beneficiaries can benefit from sending jobs to coded servers and systematic servers of the helpers.  In contrast, the helpers only need their own systematic servers.
Formally, consider the index $\bottleneck$ defined as
\begin{equation}\label{eq:bottleneck}
    \bottleneck = \max \left\{j : \left(j\sum_{i = 1}^j \allocation_i\right) < 1\right\}.
\end{equation}
Then we call job type in $\{1,2,\dots,\bottleneck\}$ the beneficiaries, and job types in $\{\bottleneck+1,\dots,k\}$ the helpers.
Note that in the special case where $k=2$, the index $\bottleneck=1$, and thus job type 1 is the beneficiary and job type 2 is the helper.
In another special case where $\allocation_i = 1/k$ for all $i$, the index $\bottleneck$ is around $\sqrt{k}$.

The choice of $\bottleneck$ in \eqref{eq:bottleneck} has an intuitive explanation based on the stability of coded servers as follows. For any $j$, if there are $j$ beneficiaries, then their total traffic is given by $\sum_{i = 1}^j \left(\allocation_in  - \slack_i\right)$. We demonstrate in \secondversion{\Cref{prf:thm:rt_analyze_general}}{\cite{choudhury2021_coded_queueing}}that a good routing policy is as follows. Each helper assigns its jobs only to its own systematic servers. Each beneficiary assigns its jobs to its own systematic servers with probability $(1 - \Theta(\numcoded/n))$, and to a recovery set consisting of $(k - j)$ systematic servers from helpers and $j$ coded servers with the rest of the probability. Under this routing policy, the total traffic to the coded servers is $j\cdot\Theta\left(\numcoded/n\right)\left(\sum_{i = 1}^j \allocation_in  - \slack_i\right)$ which is $\Theta\left(\numcoded\left(j\sum_{i = 1}^j \allocation_i\right)\right)$. The choice of the index $\bottleneck$ in \eqref{eq:bottleneck} then intuitively ensures the stability of coded servers.

Now, to define the traffic regimes, it is sufficient to look at the orders of $\slack_1$ and $\slack_{\bottleneck + 1}$. We define the traffic regimes as follows.

\paragraph{Light regime:} $\slack_i\ge 0$ for all $i\in\{1,2, \dots, k\}$, and \[\slack_i = \omega\left(\sqrt{n\numcoded}\right),\text{ for all } i\in\{1,2, \dots k\}.\]

\paragraph{Inner-heavy regime:} $\slack_i\ge 0,\text{ for all } i\in\{1,2, \dots ,k\}$, and 
\begin{gather*}
\slack_1 = o\left(\sqrt{n\numcoded}\right), \slack_1 = \Omega\left(\numcoded\right), \slack_{\bottleneck + 1} = \omega\left(\slack_1\right).
\end{gather*}
\paragraph{Outer-heavy regime:} $\slack_i \geq 0$, for all $i\in\{1,2, \dots, k\}$, and
\begin{gather*}
    \slack_1 = o\left(\numcoded\right), \slack_{\bottleneck + 1} = \omega\left(\numcoded\right).
\end{gather*}

The \emph{coded-unstable} and \emph{uncoded-unstable} regimes are defined in the same way as those given in \Cref{sec:regions_2}. Note that the assumption of $\numcoded = o(n)$ is necessary for defining the light, inner-heavy and outer-heavy regimes, but it is not required to define the uncoded-unstable and coded-unstable regimes. A sufficient condition for an arrival rate vector to lie inside the coded-unstable regime is given as $\slack_{\bottleneck + 1} = \omega(\numcoded)$ and $\slack_{\bottleneck + 1}\geq 0$. Likewise, a sufficient condition for an arrival rate vector to lie inside the uncoded-unstable regime is given as $|\slack_{\bottleneck}| = o\left(\numcoded\right), \slack_{\bottleneck} \leq 0, \slack_{\bottleneck + 1} \geq 0$, and $\slack_{\bottleneck + 1} = \omega\left(\numcoded\right)$. The proof of the sufficient conditions are given in \secondversion{\Cref{prf:sufficient}.}{\cite{choudhury2021_coded_queueing}.}

\subsection{Main Result}\label{sec:main_result}
We state our main result in \Cref{thm:rt_analyze_general}, which compares the response time in a coded system with that in an uncoded system in the five traffic regimes.

\begin{thm}[Response Time Comparison]
\label{thm:rt_analyze_general}
    Consider a multi-access cloud system consisting of $n$ servers and $k$ job types. Consider the minimum mean response times in a coded system and an uncoded system over all probabilistic job assigning policies, denoted by $\Ex{\responsetime_{\text{coded}}}$ and $\Ex{\responsetime_{\text{uncoded}}}$, respectively.  Then we have the following comparison in the five traffic regimes:
    \begin{align}
        \text{\emph{Light regime}: }&\left| \Ex{\responsetime_{\text{coded}}} - \Ex{\responsetime_{\text{uncoded}}}\right| = o(1);\label{eq:comp-light}\\
         \text{\emph{Inner-heavy regime}: }&\Ex{\responsetime_{\text{coded}}} \leq \Ex{\responsetime_{\text{uncoded}}} - \omega(1);\label{eq:comp-inner-heavy}\\
         \text{\emph{Outer-heavy regime}: }&\Ex{\responsetime_{\text{coded}}} = o\left(\Ex{\responsetime_{\text{uncoded}}}\right);\label{eq:comp-outer-heavy}\\
         \text{\emph{Uncoded-unstable regime}: } & \Ex{\responsetime_{\text{coded}}}<\infty,
         \Ex{\responsetime_{\text{uncoded}}}=\infty;
         \\
         \text{\emph{Coded-unstable regime}: } & \Ex{\responsetime_{\text{coded}}} = \infty,         \Ex{\responsetime_{\text{uncoded}}} < \infty.
    \end{align}
\end{thm}

\subsubsection{Proof sketch of Theorem \ref{thm:rt_analyze_general}}\label{sec:sub_proof_sktech}

In this subsection, we provide a proof sketch of \Cref{thm:rt_analyze_general}. We first analyze the mean response time of the uncoded system, and then analyze same for the coded one.

\paragraph{\textbf{Uncoded system.}} The arrival rate of jobs of type $i$ is $\arrival_i$, which is served using $\allocation_i n$ servers. One can prove that the optimal probabilistic routing policy is to serve an incoming job of type $i$ using one of the systematic server of type $i$ chosen uniformly at random. Under this optimal policy, all systematic servers of type $i$ behave like independent $M/M/1$ queues with arrival rate $\arrival_i/(\allocation_i n)$ and unit service rate. The mean response time of a type $i$ job is then  $1/\left(1 - \frac{\arrival_i}{\allocation_i n}\right) = \frac{\allocation_i n}{\slack_i}$. Therefore, the mean response time of the uncoded system is given by 
\begin{equation}
    \Ex{\responsetime_{\text{uncoded}}} = \sum_{i = 1}^k \frac{\arrival_i}{\sum_{\ell = 1}^k \arrival_\ell}\frac{\allocation_i n}{\slack_i}.
\end{equation}

\begin{figure*}[htp]
    \begin{subfigure}{0.7\linewidth}
        \includegraphics[width=0.32\linewidth]{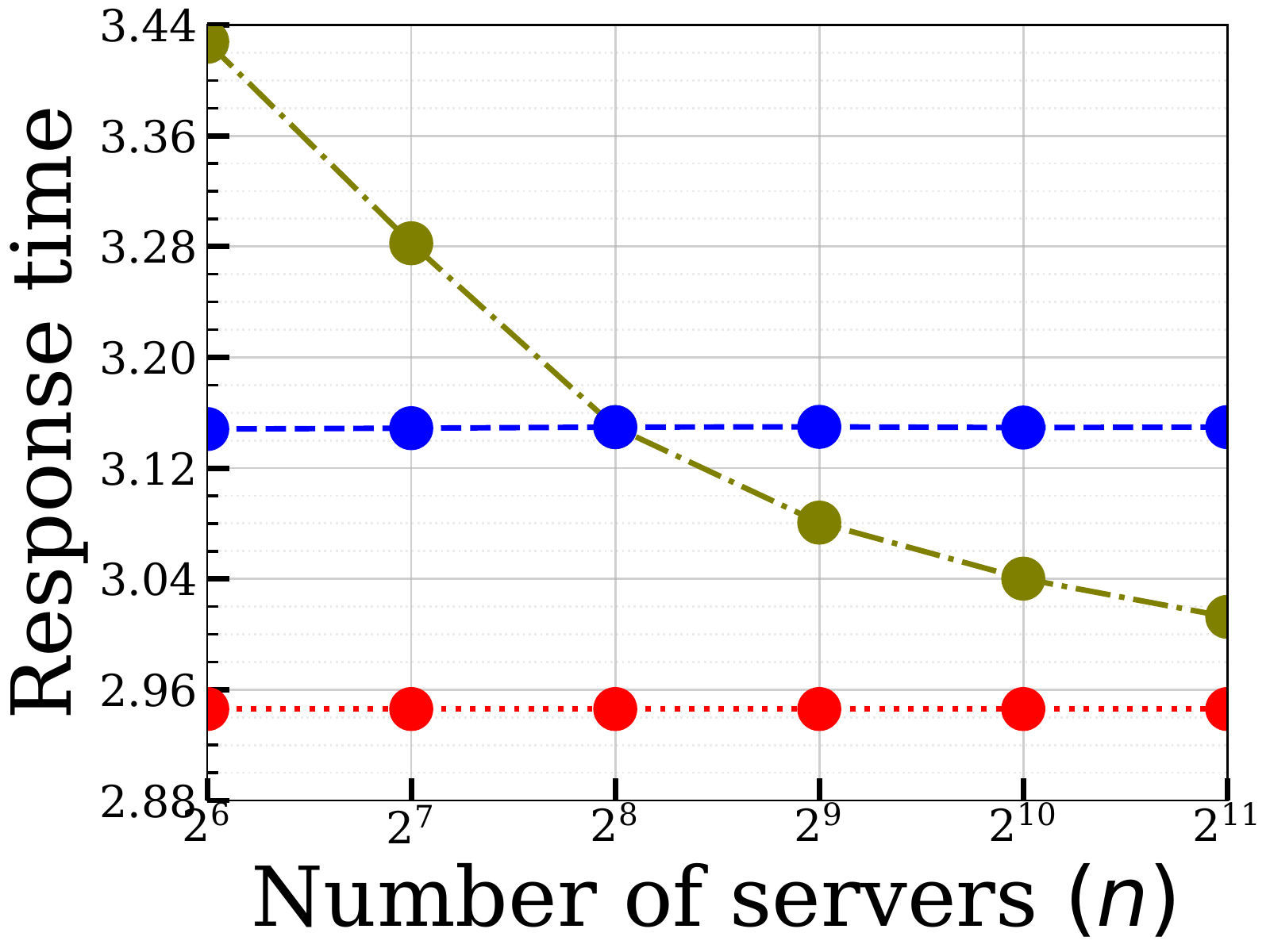}
        \includegraphics[width=0.32\linewidth]{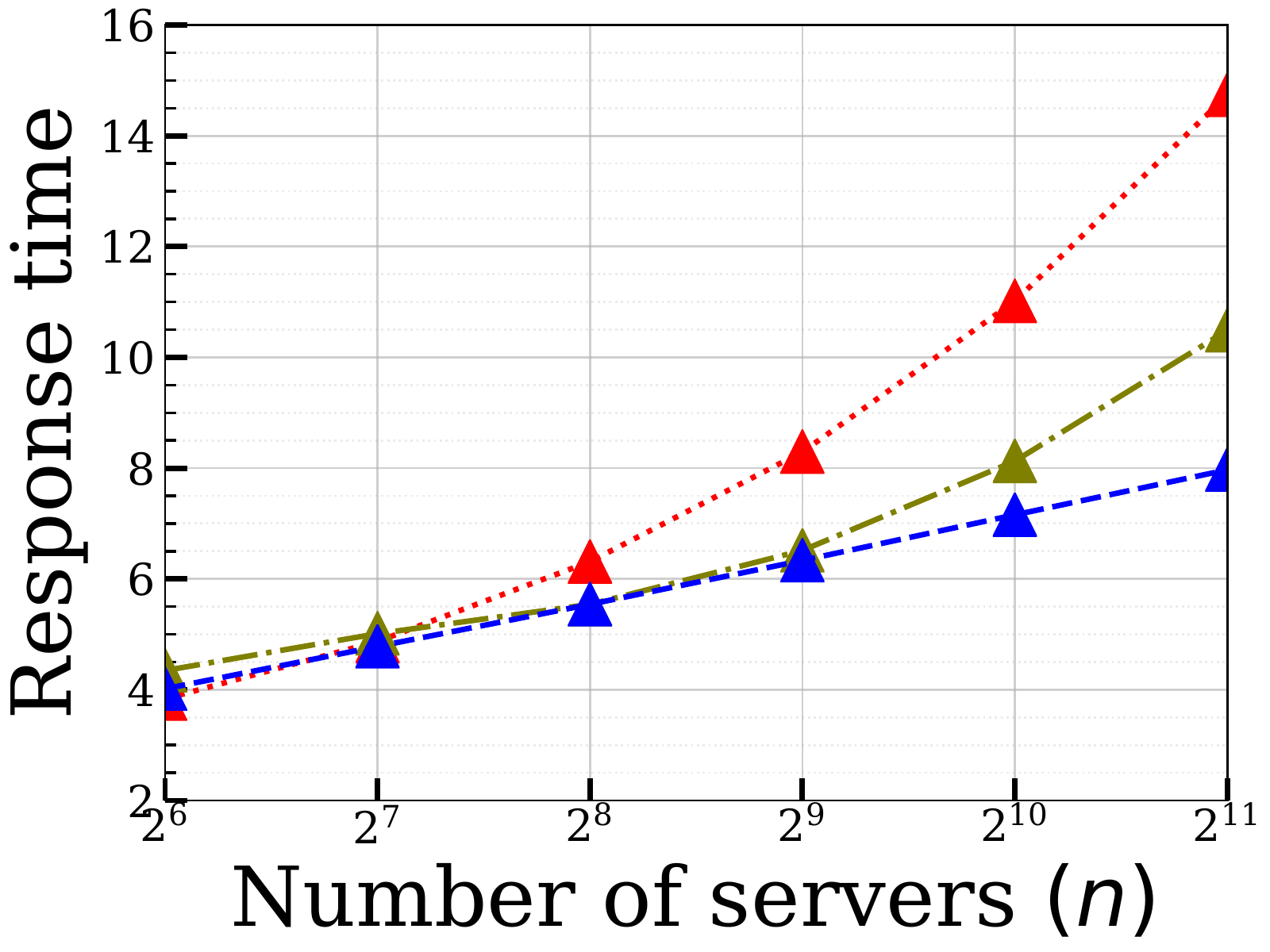}
        \includegraphics[width=0.32\linewidth]{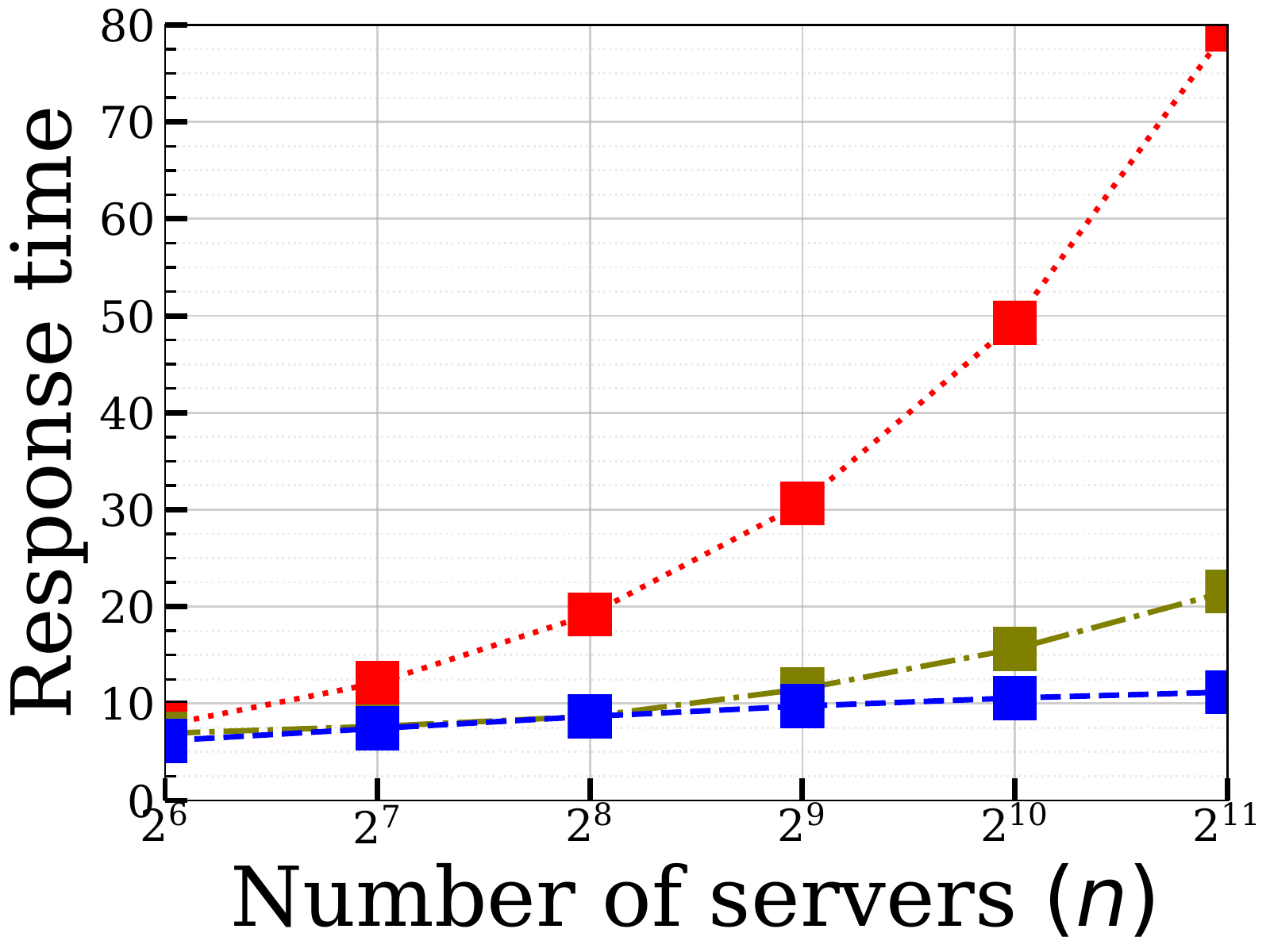}
        \caption{Response time comparison between the coded and uncoded system for $k = 2$.}
        \label{fig:rt_2file_compare_lin_o}
    \end{subfigure}
    \begin{subfigure}{.29\linewidth}
        \vspace{-10mm}
        \fbox{\includegraphics[page = 1, trim = {2mm 2mm 1mm 2mm}, clip,  width=0.99\linewidth]{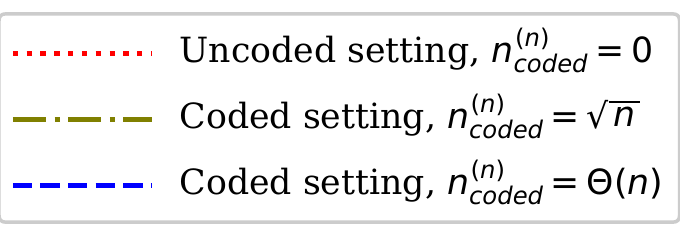}}
    \end{subfigure}
    % \vspace{10mm}
    \centering
    \begin{subfigure}{0.7\linewidth}
        \includegraphics[width=0.32\linewidth]{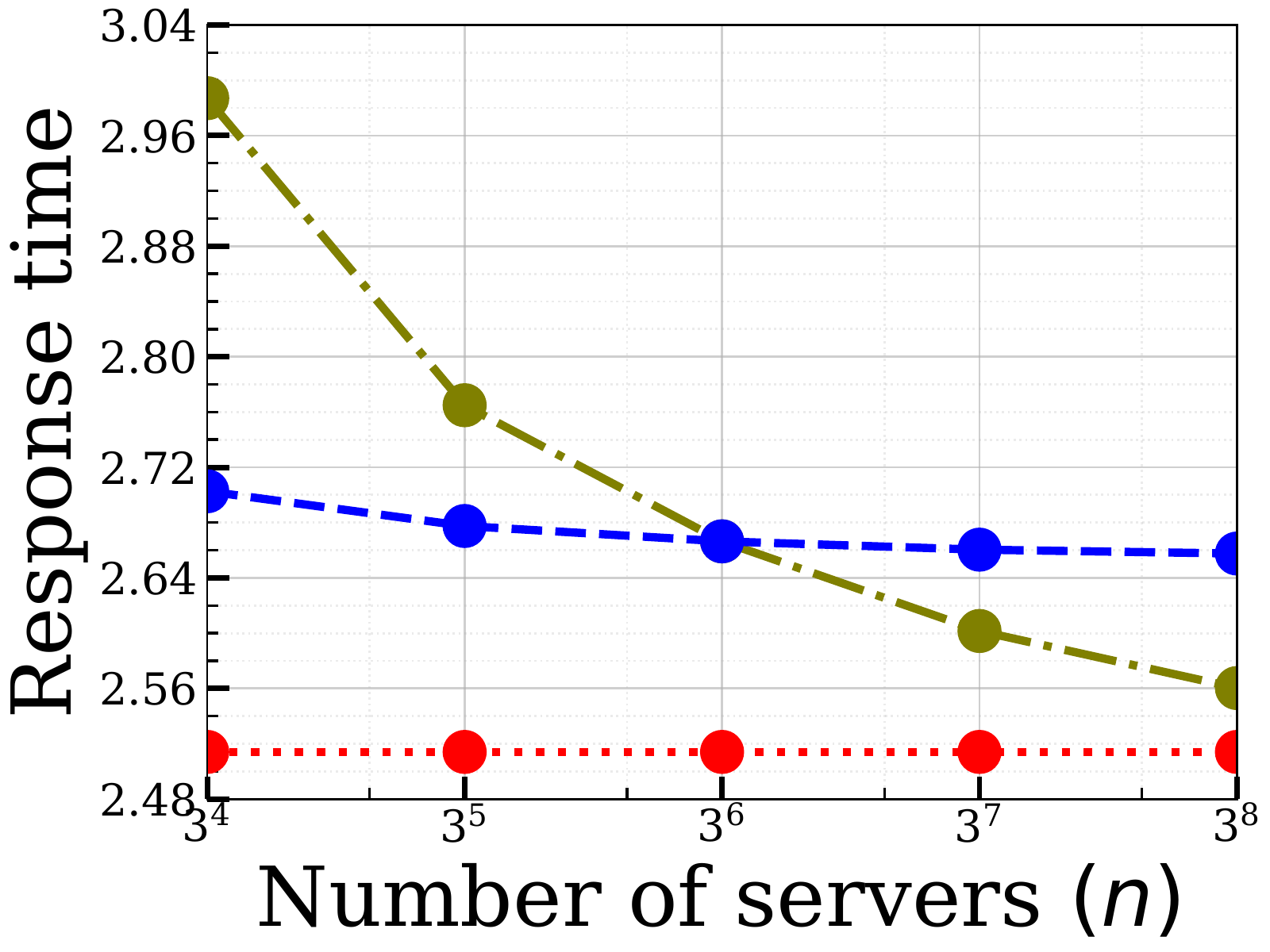}
        \includegraphics[width=0.32\linewidth]{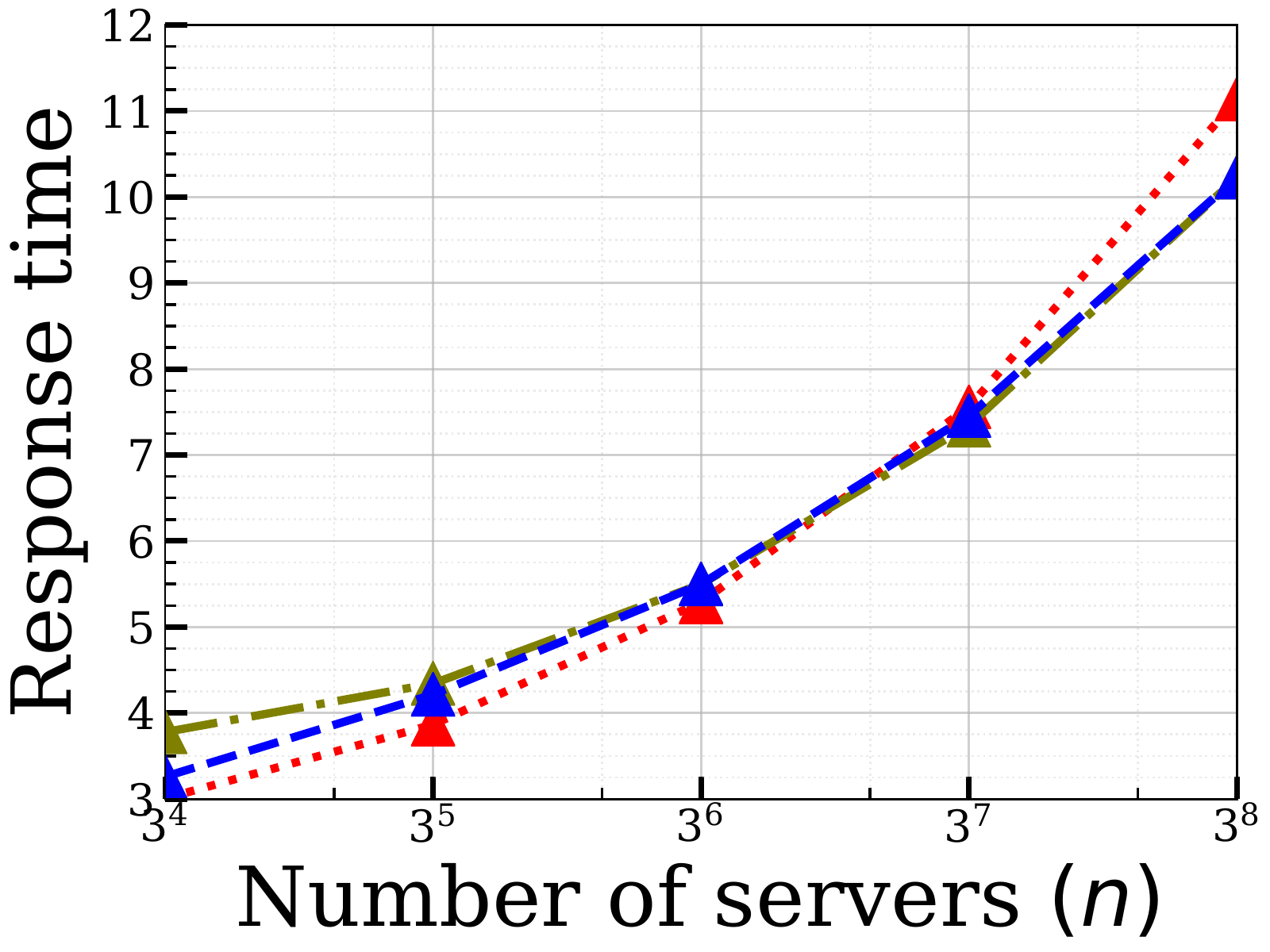}
        \includegraphics[width=0.32\linewidth]{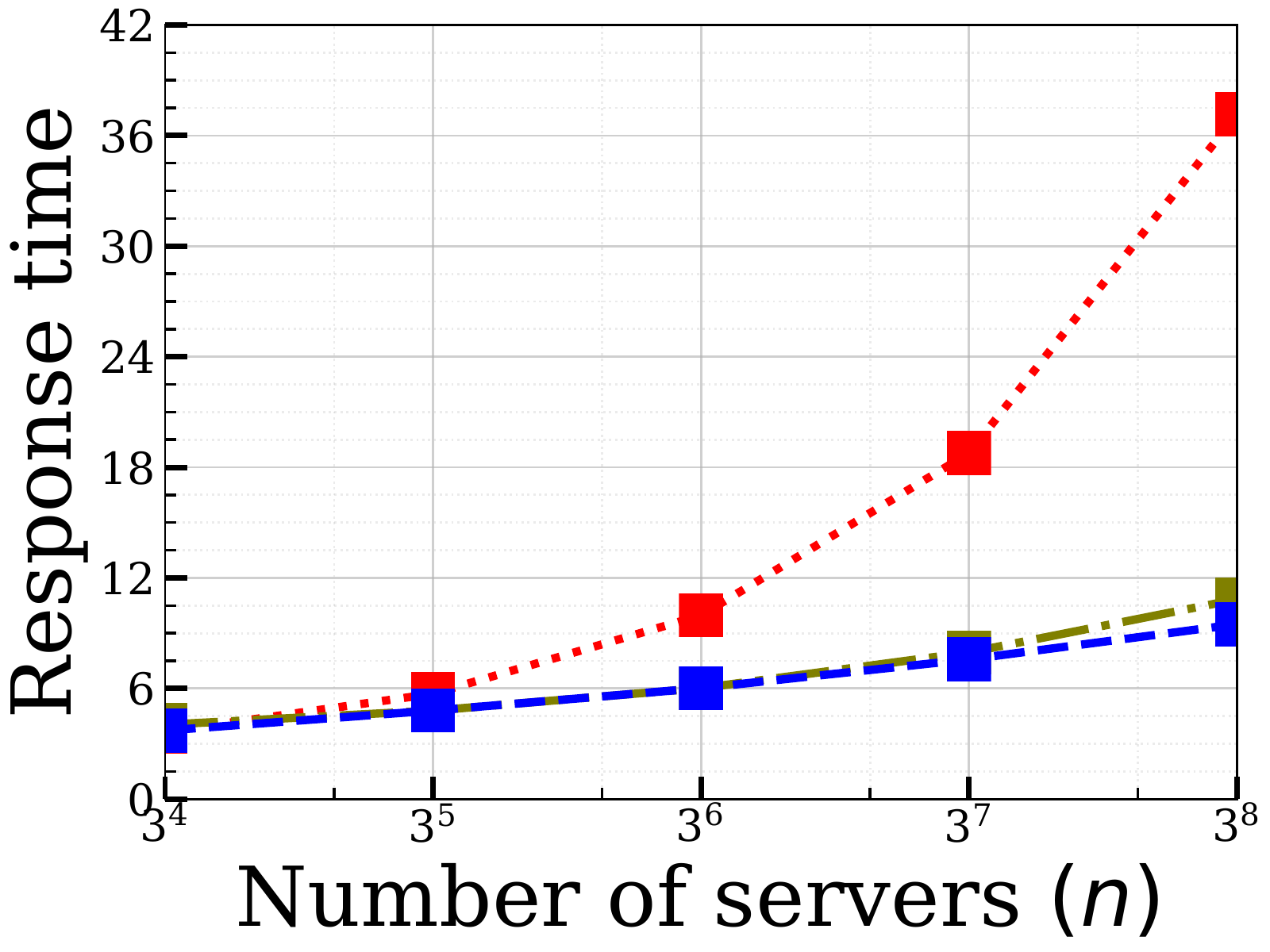}
        \caption{Response time comparison between the coded and uncoded system for $k = 3$.}
        \label{fig:rt_3file}
    \end{subfigure}
    \begin{subfigure}{0.29\linewidth}
        \vspace{-10mm}
        \hspace{-5mm}
        \includegraphics[page = 2, trim = {5mm 5mm 5mm 5mm}, clip,  width=1.15\linewidth]{finalPlots/plotmark.pdf}
    \end{subfigure}
    \centering
    \caption{Comparison of the response times of the coded and uncoded systems. The different marker types (circle, triangle, square) represent arrival rates from different traffic regimes (light, inner-heavy and outer-heavy), as shown in the bottom-right illustration. The empirical standard error of the plotted points is $O(10^{-3})$. The coded servers significantly improve the system's performance in the inner-heavy and outer-heavy regimes. The coded and uncoded systems have similar response times in the light regime. 
    }
    
    \label{fig:rt_fixed_ar}
\end{figure*}

\paragraph{\textbf{Coded system}} For the coded system, the routing probability is the key component deciding the mean response time. To better understand job routing policies in the coded system, we first give the \Cref{lem:1-p1} we observe for any stabilizing policy.

\begin{propty}\label{lem:1-p1}
    Consider any routing policy and let $\probab_{i0}$ be the total probability of sending a job of type $i$ to a systematic server of type $i$. Then, the policy can stabilize the system only if
    \begin{equation}
        1 - \probab_{i0} = O\left(\frac{\numcoded}{n}\right), \text{for all } i.
    \end{equation}
\end{propty}

\Cref{lem:1-p1} states that the fraction of traffic that any job type can divert away from its own systematic servers is limited to a $O\left(\numcoded/n\right)$ fraction. This provides a lower bound on the traffic that has to be served by systematic servers, which further leads to a lower bound on the mean response time. We use this lower bound in the analysis of the light regime. Moreover, in the heavy regimes where traffic is more skewed, we need to divert the traffic of heavily loaded job types as much as we can.  In this case, \Cref{lem:1-p1} also provides a guideline for choosing a good job routing policy. The analysis of the individual regimes then proceeds as follows. 

\begin{enumerate}[wide, labelwidth=!, labelindent=0pt]
    \item \textit{Light regime}: To prove \eqref{eq:comp-light}, it suffices to show a lower bound, 
    \begin{equation}\label{eq:comp-light_lower}
        \Ex{\responsetime_{\text{coded}}} \geq \Ex{\responsetime_{\text{uncoded}}} + o(1),
    \end{equation}
    and an upper bound given by 
    \begin{equation}\label{eq:comp-light-upper}
        \Ex{\responsetime_{\text{coded}}} \leq \Ex{\responsetime_{\text{uncoded}}} + o(1).
    \end{equation}
    As mentioned earlier, the lower bound is proved using \Cref{lem:1-p1}.  To show the upper bound, note that it suffices to focus on a particular routing policy and show that its mean response satisfies \eqref{eq:comp-light-upper}. In the light regime, the slack capacity for every job type is large enough.  Therefore, we consider the routing policy that assigns every job to its own systematic server. Computing the corresponding mean response time verifies \eqref{eq:comp-light-upper}.
    \item \textit{Heavy regimes}: The analysis of the inner-heavy and outer-heavy regimes follow the same structure. In these regimes, the beneficiaries experience heavy traffic and have small slack capacities. To reduce the mean response times, we divert the traffic of beneficiaries from their systematic servers to the recovery sets that utilize coded servers as much as possible.

    In fact, we consider the following routing policy. For some appropriate index $\bottlenecktwo \leq \bottleneck$, we choose the routing probability $\probab_{i0} = 1$ for all $i > \bottlenecktwo$. For any $i \leq \bottlenecktwo$, we use the routing probability $\probab_{i0} = 1 - v\numcoded/n$ and $\probab_{i\bottlenecktwo} = v\numcoded/n$, for some constant $v$.

    For any $i > \bottlenecktwo$, if the routing option is chosen corresponding to the probability $\probab_{i\bottlenecktwo}$, then a recovery set consisting of $\bottlenecktwo$ coded servers and $(k - \bottlenecktwo)$ systematic servers of type $\bottlenecktwo + 1, \dots, k$ respectively is chosen uniformly at random from all recovery sets satisfying the property. For any $i$, if the routing option is chosen corresponding to the probability $\probab_{i0}$, then a systematic server of type $i$ is chosen uniformly at random. Upper-bounding the mean response time for this policy gives the upper bounds in \eqref{eq:comp-inner-heavy} and \eqref{eq:comp-outer-heavy}
\end{enumerate}

\section{Simulation Results}\label{sec:simulations}
In this section, we present our simulation results to demonstrate the performance comparison between the uncoded and coded system under various traffic settings. In Section~\ref{sec:simulations_far}, we focus on arrival rates that are time-invariant. Our main goal is to demonstrate the performance comparison given in \Cref{thm:rt_analyze_general}, but we have also investigated the choice of the $\numcoded$ not covered in \Cref{thm:rt_analyze_general}. In \Cref{sec:simulations_var}, we consider arrival rates that are time-varying, with a traffic pattern commonly observed in practical systems. \secondversion{A detailed description of the simulation setups are provided in \Cref{sec:app_simulation_setup}.}{}

Before we get into the simulation settings, we first describe the routing policy we use in the simulations for the coded system. 

\begin{figure*}[thb]
    \begin{subfigure}{0.31\linewidth}
        \includegraphics[width=1\linewidth]{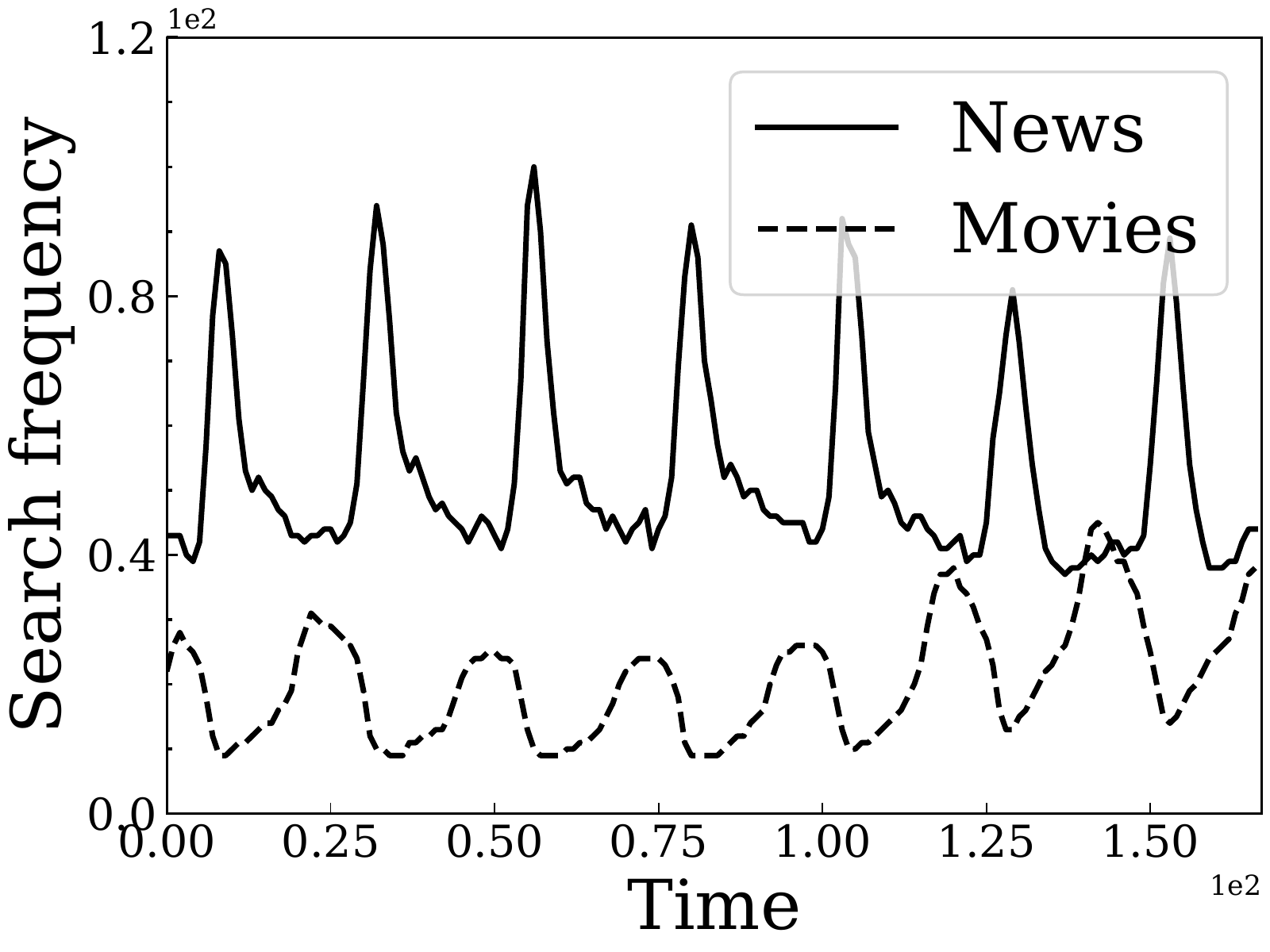}
        \caption{Search frequency on Google}
        \label{fig:trend_google}
    \end{subfigure}
    \begin{subfigure}{0.31\linewidth}
        \includegraphics[width=1\linewidth]{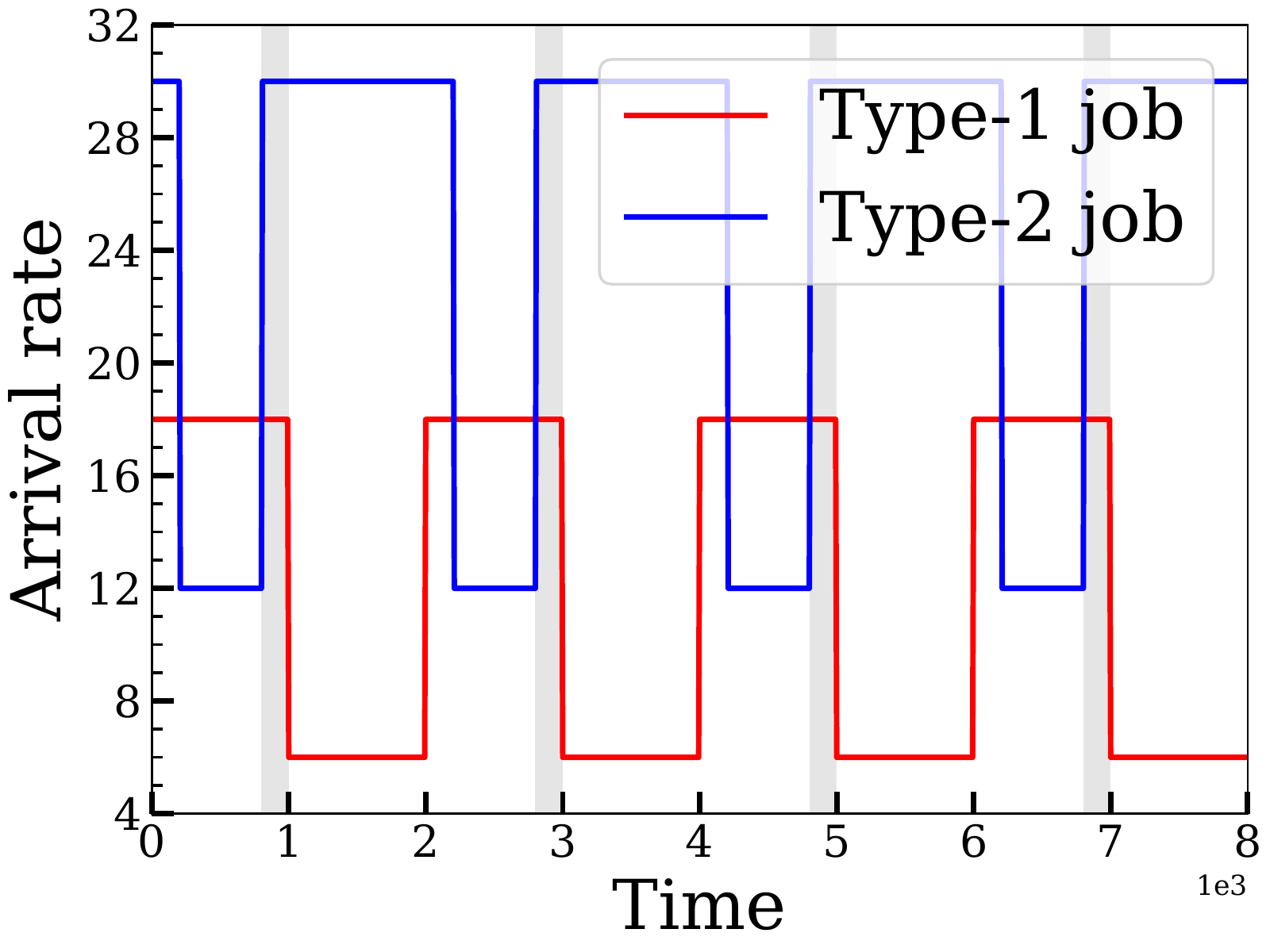}
        \caption{Arrival rates used in simulation}
        \label{fig:pulse_ar}
    \end{subfigure}
    \begin{subfigure}{0.31\linewidth}
        \includegraphics[width=1\linewidth]{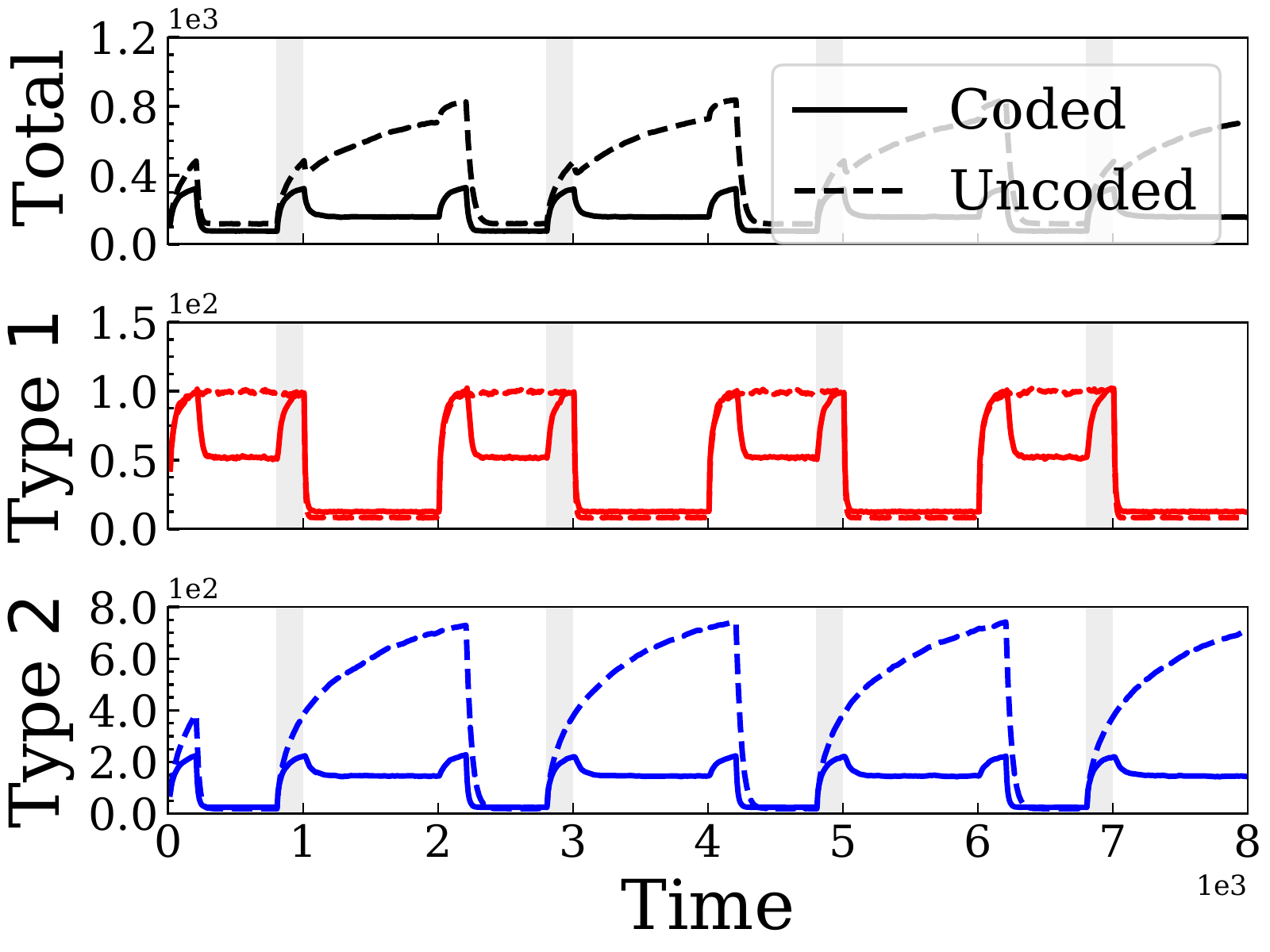}
        \caption{Number of jobs in the system}
        \label{fig:pulse_nj}
    \end{subfigure}
    \caption{
    A comparison of the coded and uncoded systems for time-varying traffic in terms of total job in the system. The leftmost picture shows the search frequency of the words \textit{news} and \textit{movies} on Google, demonstrating a real-world example of negatively correlated traffic. The central figure shows the arrival rate vector as a function of time for a simplified system with two job types that has a negative correlation in traffic. The rightmost figure shows the number of jobs in the system. The presence of coded servers helps reduce the load from the heavier loaded systematic servers, making the coded system enjoy a lesser total job in the system on average. }
    \label{fig:pulse}
\end{figure*}
\paragraph{\textbf{Pseudo-optimal routing policy}} Each queue behaves like a $M/M/1$ queue; hence the response time of each task is an exponentially distributed random variable. However, finding the optimal routing policy is non-trivial since the response time of a job is the maximum of the response time of its tasks, and the queues at each server are not independent. Moreover, the routing policy discussed in \Cref{sec:sub_proof_sktech} does not perform well empirically for smaller values of $n$ even though it works well asymptotically.  

The difficulty in obtaining the optimal routing policy is the dependence among queues. We derive a policy that we call the \emph{pseudo-optimal} routing policy by treating the queues as if they were independent. This approximation is based on the commonly observed phenomenon that queues are asymptotically independent in large systems \cite{weina_asymp_inde}. With the independence assumption, one can calculate a job's response time as the expectation of maximum of independent exponential random variables is known. The pseudo-optimal routing policy is then the policy that minimizes the approximated mean response time. In our simulations, we find the pseudo-optimal routing policy numerically using Scipy Optimization libraries.

\subsection{Time-invariant Arrival Rates }\label{sec:simulations_far}

In this subsection, we experimentally demonstrate the performance comparison between a coded system and an uncoded system. We provide simulations for systems with $2$ and $3$ job types. For these simulations, apart from using $\numcoded = o(n)$, we also consider $\numcoded = \Theta(n)$ and show the effect of large number of coded servers. We simulate until $10^8$ jobs leave the system and average it over $50$ runs to calculate the mean response time.  Finally, based on our simulation results, we provide intuitions on how our main result would change for the case of $\numcoded = \Theta(n)$.
\paragraph{\textbf{Simulation for systems with $2$ job types}} 
We consider $n = 2^m$ servers, where we vary $m\in\{6, 7, \cdots, 11\}$. For the uncoded system, we calculate the response time theoretically. For the coded system, we consider two cases of $\numcoded = o(n)$ and $\numcoded = \Theta(n)$. 

\Cref{fig:rt_2file_compare_lin_o} shows numerical comparison between mean response of the coded and uncoded system for $k=2$. The results for the case of $\numcoded = o(n)$ resembles the theoretical results provided in \Cref{thm:rt_analyze_general}. In the light regime, the coded systems with $o(n)$ coded servers perform similar to the uncoded system with a diminishing performance gap as $n$ increases. In the inner-heavy and outer-heavy regimes, the coded system with $o(n)$ servers outperforms the uncoded system and the gap increases as $n$ increases. Moreover, the performance gap between the coded and uncoded systems increases with the skewness in arrival rate, i.e., the coded system performs significantly better in the outer-heavy regime.

However, as illustrated in \Cref{fig:rt_2file_compare_lin_o}, the results are slightly different when the number of coded servers increases as $\numcoded = \Theta(n)$. In the light regime, the coded system with $\Theta(n)$ coded server performs worse than the uncoded system. The redundancy added by $\Theta(n)$ coded servers worsens the system. However, the performance gap does not change much as $n$ increases as the traffic is light enough. The cost of redundancy is not substantial, and the coded system is worse only by an $\Theta(1)$ term. Compared to the light regime, the coded system with $\Theta(n)$ servers performs considerably better than both uncoded and coded systems with $o(n)$ servers in the inner-heavy and outer-heavy regimes. Because of the higher skew, the coded system with $\Theta(n)$ coded servers allows more uniform load balancing, thus greatly improving the performance. 

\paragraph{\textbf{Simulation for systems with $3$ job types}} We also perform experiments for system with three types of jobs. We consider $n = 3^m$ servers, where we vary $m\in \{4, 5, \cdots, 8\}$.

\Cref{fig:rt_3file} shows the response time comparison of the coded and uncoded system for this simulation setup. For the light and outer-heavy regime, the trends in \Cref{fig:rt_3file} is similar to the simulation results for the two-job type system and hence follows a similar reasoning. However, for the inner-heavy regime, the trends are slightly different. The coded system outperforms the uncoded system asymptotically; however, the difference is not as significant as seen in the simulation with two job types. One plausible reason is that the system has two beneficiaries and only one helper based on our arrival rate choice and allocation of servers. Hence, a slight skew in the arrival rate vector is insufficient to reduce the beneficiaries' load. However, there is enough skew in the outer-heavy region such that the coded system outperforms the uncoded system regardless of the number of coded servers. 

Based on the simulations for systems with two or three job types, we conjecture that for $\numcoded = \Theta(n)$, the inner-heavy and outer-heavy regimes would merge into a single heavy regime where the coded system would outperform the uncoded system. The light regime would also change, and instead of $o(1)$, the mean response time's of the coded and uncoded system can differ by $\Theta(1)$.

\subsection{Time-Varying Arrival Rates}\label{sec:simulations_var}
In practical systems such as Google search, the traffic of a job type often varies periodically. For example, as shown in \Cref{fig:trend_google}, search for news in Google \cite{googletrends} peaks during early hours, while the search for movies peaks during the night. To simulate such traffic patterns, we consider a system with two job types where the arrival rate of each job type is a square wave, as illustrated in \Cref{fig:pulse_ar}. The arrival rates of the two job types have a similar period but are negatively correlated, i.e., if one job experiences higher traffic, the other job type should experience lesser traffic. We consider a total of $n=60$ servers. In the coded system, $7$ of these servers are coded servers. The remaining servers are distributed in the same proportion to the two job types for the coded and uncoded systems.

\Cref{fig:pulse_nj} shows a comparison between the mean number of jobs in the coded and uncoded systems. The top plot shows the total number of jobs in the system as a function of time, while the second and the third plot shows the number of jobs of type $1$ and $2$, respectively. When a job type is experiencing low traffic, the number of jobs of that type in the uncoded system is slightly less than that in the coded system. However, when a job type is experiencing heavy traffic, the number of jobs in the uncoded system is significantly higher than the coded system. This is because coding provides a load balancing effect where lightly loaded servers can be used to serve heavier traffic for job types. 

\section{Conclusion}\label{sec:conclusion}
This paper proposes the use of erasure-coded servers to handle traffic variations in heterogeneous jobs. We show that adding a few erasure coded servers significantly expands the capacity region of the coded system thereby improving the stability. We also compare the latency of the coded and uncoded systems and show that the coded system is better or at least comparable in most traffic regimes. The erasure-coded servers also improve the system's flexibility as the system can quickly adapt to changes in traffic, especially when the traffic is negatively correlated. Thus, at a slight cost of redundancy, our coded solution provides a general framework to improve the system's stability and latency.

There are substantial directions for future work. While our proposed coded system to handle $k$ job types and its analysis holds for any $k$, in practice, a large $k$ would be impractical because the decoding cost scales as $O(k^3)$. To handle large $k$, we could divide the servers into $r$ subsystems with $k/r$ job types in each subsystem. A large $r$ will save the decoding cost, but it loses some flexibility offered by coding. In future work, we can determine the optimal choice of $r$ and strategies to group the $k$ job types into the $r$ subsystems. Another future direction is to consider redundant replicas of a job that are sent to multiple recovery sets. A job is served when any one of the requests is served. Finally, we also plan to analyze our system for queue-length-based routing policies instead of probabilistic policies considered in this paper. Then the technical challenges in comparing the latency involve proving a complicated state-space collapse and lower bounding the mean response time.

\section*{Acknowledgments}
This work was supported in part by the NSF CCF \#2007834 and \#2045694, NSF CNS \#2007733 and \#2112471, NSF ECCS \#2145713, and a Carnegie Bosch Institute Research Award. We thank Mor Harchol-Balter and Isaac Grosof for helpful discussions. \secondversion{The authors are grateful to Mor Harchol-Balter and Isaac Grosof for helpful discussions.}{}

%\nocite{*}
\bibliographystyle{ACM-Reference-Format}
\bibliography{refs-tuhin}
\clearpage
\onecolumn
\secondversion{\appendix
\addtocontents{toc}{\protect\contentsline{chapter}{Appendix:}{}}
\section{Bachmann–Landau notation}\label{sec:landau}
\begin{defn}[Bachmann–Landau notation]\label{def:bachman_landau}
For any functions $f$ and $g$, the Bachmann–Landau notations are given below
\begin{enumerate}
    \item Small o: 
    \begin{equation}
        f = o(g)  \triangleq \lim_{n\rightarrow \infty} \frac{f(n)}{g(n)} = 0.
    \end{equation}
    \item Big O: 
    \begin{equation}
        f = O(g)  \triangleq \lim\sup_{n\rightarrow \infty} \frac{f(n)}{g(n)} < \infty.
    \end{equation}
    \item Small omega: 
    \begin{equation}
        f = \omega(g)  \triangleq \lim_{n\rightarrow \infty} \left|\frac{f(n)}{g(n)}\right| = \infty.
    \end{equation}
    \item Big Theta:
    \begin{equation}
        f = \Theta(g)  \triangleq \exists k_1, k_2 > 0 \mbox{ s.t. } k_1 \leq \lim_{n\rightarrow \infty} \frac{f(n)}{g(n)} \leq k_2.
    \end{equation}
\end{enumerate}
\end{defn}
\section{Proof of Service-Capacity region}\label{prf:capacity_region}
In this section, we prove the service capacity region of system given in \eqref{eq:stability_condition_coded}. We then determine the service capacity region for the special case of $k=2$. The proof for the special case is taken from \cite{coded_queueing_mehmet, coded_queueing_mehmet_journal} and is included for completion.
\subsection{Proof of Equation \ref{eq:stability_condition_coded}}
\begin{proof}\label{prf:service_capacity_region}
    Recall the notations given in \eqref{eq:stability_condition_coded}. The residual capacity for a job of type $i$ is given by $r_i = \allocation_i(n - n_{\text{coded}}) - \lambda_i$. Also, $r_{i}^{+}$ and $r_{i}^{-}$ are defined as $\max\{r_i, 0\}$ and $-\min\{r_i, 0\}$, respectively. We also assume $r_1 \leq r_2\leq r_3 \leq \dots \leq r_k$. We need to prove that the system is stable if and only if
    \begin{equation}\label{eq:stability_condition_coded_restate}
        \min_{k_0 \in \{1, \dots ,k\}}\left\{\frac{n_{\text{coded}} + \sum_{i = 1}^{k_0} r_{i}^+}{k_0}\right\} \geq \sum_{i = 1}^k r_{i}^{-}.
    \end{equation}

    The proof consists of two parts. 
    \begin{enumerate}
        \item \emph{Proof of \say{if} statement}: Using the water-filling argument, we prove that if \eqref{eq:stability_condition_coded_restate} is true, the system is stable.
        \item \emph{Proof of  \say{only if} statement}: We prove that if \eqref{eq:stability_condition_coded_restate} is not true, then the system is unstable. To prove this, we prove that if an arrival rate vector lies in the service capacity region, it must satisfy \eqref{eq:stability_condition_coded_restate}.
    \end{enumerate}

    \paragraph{\textbf{Proof of if statement}} We first fill the capacity of the systematic servers, i.e., serve a job using only its dedicated systematic server to ensure minimum redundancy in the system. Once the systematic servers reach full capacity, we gradually fill the coded servers.
    
    Based on the notations, the volume of unserved jobs after we fill the systematic servers is $\sum_{i = 1}^{k} r_{i}^{-}$. The remaining capacity of systematic servers of type $i$ is $r_i^{+}$ and the capacity of the coded servers is $n_{\text{coded}}$. The system is stable if we can serve the remaining unserved jobs using a coded combination. We prove that using the coded combination, we can serve a volume of $\min_{k_0 \in \{1, \dots ,k\}}\left\{\frac{n_{\text{coded}} + \sum_{i = 1}^{k_0} r_{i}^+}{k_0}\right\}$.
    
    If $r_{1} \geq 0$, then \eqref{eq:stability_condition_coded_restate} is satisfied because each term on the left hand side is greater than or equal to zero and the right hand side is exactly zero.  Hence, for the remaining proof, we assume that $r_{1} < 0$. Let $m$ be the minimum index such that $r_m > 0$. 
    
    Initially, we start filling the systematic servers $m, m + 1, \dots, k$. If we add $v$ volume of water in the systematic servers, we add $(m - 1)v$ volume of water in the coded server. Either the coded servers or the systematic server of type $m$ will fill up. If the coded servers reach the maximum capacity, we stop. Else, we start filling the systematic servers of type $m + 1, \dots, k$. For the coded servers, we add water at a $m$ time faster rate. We continue this way until the coded servers fill up. If all systematic servers fill up, we then start filling the coded servers by a multiple of $k$. We now prove that we can add $\min_{k_0 \in \{1, \dots ,k\}}\left\{\frac{n_{\text{coded}} + \sum_{i = 1}^{k_0} r_{i}^+}{k_0}\right\}$ volume of water using this method. 
    
    Observe that after the first step, if the coded servers fill up, then the total volume of water added is $n_{\text{coded}}/(m - 1)$. If not, the remaining capacity of the coded servers is $(n_{\text{coded}} - (m - 1)r_m^+)$. In the next step, if the coded servers fill up, the total volume  water added is $r_m^+  + (n_{\text{coded}} - (m - 1)r_m^+)/m = (n_{\text{coded}}  + r_m^+)/m$. If not, the remaining capacity of the coded servers is $(n_{\text{coded}} - (m - 1)r_m^+ - m(r_{m +1}^+ -  r_m^+) = (n_{\text{coded}} + r_m^+ - mr_{m +1}^+)$. If the coded servers fill up in the third step, the total volume of added $r_{m + 1}^+ + (n_{\text{coded}} + r_m^+ - mr_{m +1}^+)/m + 1 = (n_{\text{coded}} + r_m^+  + r_{m +1}^+)/(m + 1)$. Following a similar argument, if the coded servers fill up at the $n$-th step, then the total volume of water added is $(n_{\text{coded}} + r_m^+  + r_{m +1}^+ + \dots r_{m + n - 2})/(m + n - 2)$. Moreover, at the end of step $p<n$, the remaining capacity of coded servers is $(n_{\text{coded}} + r_m^+  + r_{m +1}^+ + \dots  + r_{m + p - 2} - (m + p - 2)r_{m + p - 1})/(m + p - 1)$ which is positive. This implies, for all $p < n$,
    \begin{equation}
        \frac{n_{\text{coded}} + \sum_{j = m}^{m + p - 1}r_j^+}{m + p - 1} < \frac{n_{\text{coded}} + \sum_{j = m}^{m + p - 2}r_j^+}{m + p - 2}.
    \end{equation}
    Moreover, since the coded servers fill up at the $n$-th step, for all $p > n$,
    \begin{equation}
        \frac{n_{\text{coded}} + \sum_{j = m}^{m + p - 1}r_j^+}{m + p - 1} > \frac{n_{\text{coded}} + \sum_{j = m}^{m + p - 2}r_j^+}{m + p - 2}.
    \end{equation}
    This proves that the volume of water added is exactly
    \begin{equation}
        \min_{k_0 \in \{1, \dots ,k\}}\left\{\frac{n_{\text{coded}} + \sum_{i = 1}^{k_0} r_{i}^+}{k_0}\right\} = \min_{k_0 \in \{1, \dots ,k\}}\left\{\frac{n_{\text{coded}} + \sum_{i = m}^{k_0} r_{i}^+}{k_0}\right\}=  \frac{n_{\text{coded}} + r_m^+  + r_{m +1}^+ + \dots r_{m + n - 2}}{m + n - 2}.
    \end{equation}
    
    We also show that the maximum volume of water filled using $r_i^+$ capacity of systematic servers of type $i$, for all $i$ and $n_{\text{coded}}$ coded capacity cannot exceed $\min_{k_0 \in \{1, \dots ,k\}}\left\{\frac{n_{\text{coded}} + \sum_{i = 1}^{k_0} r_{i}^+}{k_0}\right\}$. To prove this, assume $V_{\max}$ is the maximum volume of water that can be added. Then the total volume of water added in servers of type $1, 2, \dots, k_0$ and coded servers is at least $k_0 V_{\max}$. This is because anytime one of the systematic servers in $\{1, 2, \dots, k_0\}$ is not used, the rate at which water is added in the coded servers increases by $1$, keeping the net rate of water addition at least $k_0$. But since total volume of systematic servers of type $i$ for all $i\in \{1, 2, \dots, k_0\}$ and coded servers is $(n_{\text{coded}} + \sum_{i = 1}^{k_0}r_i^+)$, we have $k_0 V_{\max} \leq (n_{\text{coded}} + \sum_{i = 1}^{k_0}r_i^+)$. Hence, for any $k_0$, $V_{\max} \leq \frac{n_{\text{coded}} + \sum_{i = 1}^{k_0}r_i^+}{k_0}$, which completes the proof.
    \paragraph{\textbf{Proof of only if statement}} We now prove that if an arrival rate vector lies in the service capacity region, it must satisfy \eqref{eq:stability_condition_coded_restate}. Assume that the arrival rate vector $\bm{\lambda} = (\lambda_1, \dots, \lambda_k)$ lies in the interior of the service capacity region. Using the definition of service capacity region in \Cref{lem:capacity_region_coded}, we know there exists $\lambda_{ij} \geq 0$ for all $i \in \{1, \dots, k\}$, $j \in \{1, 2, \dots,  |\recovery_i|\}$ satisfying
    \begin{equation}\label{eq:stability_coded_1_restate}
        \lambda_i = \sum_{j =1}^{ \mid\recovery_i \mid} \lambda_{ij}, \forall i\in\{1, \dots, k\},
    \end{equation}
    \begin{equation}\label{eq:stability_coded_2_restate}
        \sum_{i = 1}^k\sum_{j: \ell\in \recovery_{ij}} \lambda_{ij} \leq 1, \forall \ell\in \{1, \dots, n\}
    \end{equation}
    We can assume that there exists no $i, i'$ s.t. there exists some $j\in \{1, \dots, |\recovery_i|\}$ and $j'\in \{1, \dots, |\recovery_{i'}|\}$, for which $\lambda_{ij} > 0$,  $\lambda_{i'j'} > 0$,  $\recovery_{ij}$ contains a systematic server of type $i'$ and $|\recovery_{i'j'}| > 1$. If there exists any such $i, i'$, then we can use the following operation multiple times such that the condition becomes true. 
    \begin{enumerate}
        \item If $\lambda_{ij} < \lambda_{i'j'}$: Change $\lambda_{ij} = 0$, $\lambda_{i'j'} = \lambda_{i'j'} - \lambda_{ij}$. Let $\ell$ be the systematic server of type $i'$ that lies is in $\recovery_{ij}$. Change $\lambda_{i'm}$ to $\lambda_{i'm} + \lambda_{ij}$ where $\recovery_{i'm} = \{\ell\}$. If $\recovery_{i'j'}$ contains a systematic server of type $i$, say $n$ then change $\lambda_{ik} = \lambda_{ik} + \lambda_{ij}$ where $\recovery_{ik} = \{n\}$, else change $\lambda_{ik} = \lambda_{ik} + \lambda_{ij}$, where $\recovery_{ik} = \recovery_{i'j'}$.
        \item If $\lambda_{ij} > \lambda_{i'j'}$: Change $\lambda_{i'j'} = 0$, $\lambda_{ij} = \lambda_{ij} - \lambda_{i'j'}$, $\lambda_{i'k} = \lambda_{i'k} + \lambda_{i'j'}$ where $\recovery_{i'k}$ contains only the systematic server of type $i'$ that is in $\recovery_{ij}$. If $\recovery_{i'j'}$ contains a systematic server of type $i$ say $\ell$, change $\lambda_{im} = \lambda_{im} + \lambda_{i'j'}$, where $\recovery_{im} = \{\ell\}$. Else, change $\lambda_{im} = \lambda_{im} + \lambda_{i'j'}$, where $\recovery_{im} = \recovery_{i'j'}$.
    \end{enumerate}
    Note, the above operation does not increase the net arrival rate of tasks to any server. Hence, the conditions in \eqref{eq:stability_coded_1_restate} and \eqref{eq:stability_coded_2_restate} are still valid. A consequence of the above assumption is that if a job of type $i$ uses a coded combination, it can only use a systematic server of type $j$ if the job of type $j$ only uses its systematic server.

    Let $\Gamma$ be the set of indices $i$'s such that jobs of type $i$ only use its systematic servers to serve its job. For all $i$, let $\hat{r}_i^{+}$ be the remaining service capacity of systematic servers of type $i$. Note, for all $i\in \Gamma^C$, $\hat{r}_i^{+} = 0$, since the remaining capacity cannot used. Similarly, for all $i$, let $\hat{r}_i^{-}$ be the volume of the job of type $i$ that is served using a coded combination. Using the fact that the arrival rate vector is in the service capacity region and arguments in the proof of the \emph{Forward} part, we know that 
    \begin{equation}
        \min_{k_0 \in \{1, \dots ,k\}}\left\{\frac{n_{\text{coded}} + \sum_{i = 1}^{k_0} \hat{r}_{i}^+}{k_0}\right\} \geq \sum_{i = 1}^k \hat{r}_{i}^{-}.
    \end{equation}
    Since the arrival rate vector may not fill the systematic servers, $\hat{r}_i^{-} \geq r_i^{-}$. Also, since for all $i\in \Gamma^C$, $\hat{r}_i^{+} = 0$, we have for all $i$, $r_i^{+} \geq \hat{r}_i^{+}$. Hence, 
        \begin{equation}
        \min_{k_0 \in \{1, \dots ,k\}}\left\{\frac{n_{\text{coded}} + \sum_{i = 1}^{k_0} r_{i}^+}{k_0}\right\} \geq \min_{k_0 \in \{1, \dots ,k\}}\left\{\frac{n_{\text{coded}} + \sum_{i = 1}^{k_0} \hat{r}_{i}^+}{k_0}\right\} \geq \sum_{i = 1}^k \hat{r}_{i}^{-} \geq \sum_{i = 1}^k r_{i}^{-}.
    \end{equation}
    This implies that if an arrival rate lies within the service capacity region, it must satisfy \eqref{eq:stability_condition_coded_restate}.
    
\end{proof}
\subsection{Service capacity region for $k=2$}
\begin{proof}
    \begin{enumerate}
        \item $0 \leq \lambda_1 \leq \allocation_1 n - (1 + \allocation_1)\numcoded$: The total service capacity of $\set_{1}$ servers is $\allocation_1(n - \numcoded)$. If all requests for jobs of type $1$ are sent to $\set_{1}$ servers, then remaining capacity of $\set_{1}$ servers is at least $\numcoded$. Total service capacity of $\set_{2}$ servers is $\allocation_2(n - \numcoded)$. Therefore, the maximum arrival rate that can be supported for jobs of type $2$ is $\allocation_2(n - \numcoded)$ plus the minimum of total service capacity of the coded servers and remaining service capacity of $\set_{1}$ servers. Hence, $\arrival_2 \leq \allocation_2 n  + \allocation_1\numcoded$.
        \item $\allocation_1 n - (1 + \allocation_1)\numcoded \leq \arrival_1 \leq \allocation_1(n - \numcoded)$: If all requests for jobs of type $1$ are sent to $\set_1$ servers, then the remaining capacity of $\set_{1}$ servers is $\allocation_1(n - \numcoded) - \arrival_1$. Hence, the maximum arrival rate for jobs of type $2$ that may be sent to a recovery set consisting of one coded server and one systematic server of type $1$ is $(\allocation_1(n - \numcoded) - \arrival_1)$. The remaining capacity of $\set_{\text{coded}}$ servers is $\numcoded - \allocation_1(n - \numcoded) + \arrival_1$. Hence, the maximum arrival rate request that may be sent to the recovery sets consisting of two coded servers is $(\numcoded - \allocation_1(n - \numcoded) + \arrival_1)/2$. Total service capacity of $\set_{2}$ servers is $\allocation_2(n - \numcoded)$. Hence, 
        \begin{align}
            \arrival_2 &\leq \allocation_2(n  - \numcoded) + \allocation_1(n - \numcoded) - \arrival_1 + (\numcoded - \allocation_1(n - \numcoded) + \arrival_1)/2\\
            &= (n - \numcoded)\left(1 - \frac{\allocation_1}{2}\right) + \frac{\numcoded}{2} - \frac{\arrival_1}{2}.
        \end{align}
        \item $\allocation_1(n - \numcoded) \leq \arrival_1 \leq \allocation_1 n + (1/2 - \allocation_1)\numcoded$: If the entire service capacity of $\set_1$ is used for jobs of type $1$, then the remaining required service capacity for jobs of type $1$ is $\arrival_1 - \allocation_1(n - \numcoded)$, which would be served by $\set_{\text{coded}}$ servers. Hence, the remaining service capacity of $\set_{\text{coded}}$ servers is $\numcoded - 2(\arrival_1 - \allocation_1(n - \numcoded)))$. Hence, the maximum arrival rate request possible for jobs of type $1$ is 
        \begin{equation}
            \arrival_2 \leq \allocation_2(n  - \numcoded) + \frac{\numcoded - \left(2\arrival_1 - 2\allocation_1(n - \numcoded)\right)}{2} = \frac{(2n - \numcoded)}{2} - \arrival_1.
        \end{equation}
    \end{enumerate}
    For the remaining portion, we find the service capacity region by fixing $\arrival_2$ and analyzing the range of value of $\arrival_1$. 
\end{proof}

\section{Proof of Theorem \ref{thm:rt_analyze_general}}\label{prf:thm:rt_analyze_general}
Recall, our system consists of $n$ servers that provide service to $k$ job types. The $n$ servers are divided into two parts, $\numcoded$ coded servers, and the remaining are systematic servers. $\allocation_i$ represents the fraction of systematic servers dedicated to the job of type $i$. Hence, $\numserver_i$, the number of systematic servers for the $i$-th job type is given by 
\begin{equation}
    \numserver_i = \allocation_i(n - \numcoded). 
\end{equation}
The arrival rate of job type $i$ is $\arrival_i$ and, the residual capacity for job type $i$ is  $\slack_i$, defined as
\begin{equation}
    \slack_i = \allocation_i n - \arrival_i.
\end{equation}

Let $\probab_{ij}$ denote the probability of serving a job of type $i$ using $j$ coded servers. $\Ex{\responsetime_{\text{coded}}}$ and $\Ex{\responsetime_{\text{uncoded}}}$ denotes the response time of the coded and uncoded system, respectively. Let $\psarrival_i$ denote the arrival rate of tasks to any systematic server of type $i$.  With the notations defined, we first determine the response time of the uncoded system.

In the uncoded system, queues corresponding to systematic servers of type $i$ behave like independent $M/M/1$ systems with arrival rate $(\arrival_i/\allocation_i n) = (1 - (\slack_i/\allocation_i n))$. Hence, the response time of jobs of type $i$ is $(\allocation_i n/\slack_i)$. Moreover, the fraction of jobs that belongs to type $i$ out of all incoming jobs is $(\arrival_i/\sum_{\ell = 1}^k \arrival_\ell)$. Hence, the mean response time of the uncoded system is given by, 
\begin{equation}\label{eq:mrt_2_un}
    \responsetime_{\text{uncoded}} = \sum_{i = 1}^k \frac{\arrival_i}{\sum_{\ell = 1}^k \arrival_\ell}\frac{\allocation_i n}{\slack_i}.
\end{equation}
Because of our assumption that  $\slack_1\leq \slack_2\leq \dots \leq \slack_k$, the dominant term in the response time of the uncoded system is $(n/\slack_1)$. 

In the remainder of the proof, we bound the response time of the coded system. Notice that the response time of the coded system depends on the routing probability $\probab_{ij}$'s. Hence, the optimal response time of the coded system depends on the optimal routing probability. However, because of the combinatorial nature, solving the optimal routing probability is a complicated problem. In addition, the fork-join nature of the coded system makes the queues dependent. Hence, there is no closed-form expression of the response time in terms of routing probability and arrival rate vectors. We overcome this problem by using an upper bound expression of the response time for a \textit{reasonable} routing probability vector. The proof consists of three separate parts, one for each regime.

\subsection{The Light Regime}
\begin{proof}[Proof of Theorem~\ref{thm:rt_analyze_general} in the Light Regime]
    The light regime is the set of arrival rate vectors that satisfy $\slack_i \leq \slack_j$, for all $i \leq j$ and $\slack_1 = \omega\left(\sqrt{n\numcoded}\right)$. For this region, we prove that 
    \begin{equation}\label{eq:asymp:compare:equal_2}
        \left| \responsetime_{\text{coded}} -  \responsetime_{\text{uncoded}}\right| = o(1).
    \end{equation}
    In order to prove \eqref{eq:asymp:compare:equal_2}, we lower bound and upper bound the response time of the coded system. In particular, we prove the following properties. 
    \begin{enumerate}
        \item For any routing policy, the response time of the coded system can be bounded by
        \begin{equation}\label{prf:light_2file_lower}
            \responsetime_{\text{coded}} \geq  \responsetime_{\text{uncoded}} +  o(1).
        \end{equation}
        \item There exists a routing policy for which the response time of the coded system can be bounded by
        \begin{equation}\label{prf:light_2file_upper}
            \responsetime_{\text{coded}} \leq  \responsetime_{\text{uncoded}} +  o(1).
        \end{equation}
    \end{enumerate}
    Combining \eqref{prf:light_2file_lower} and \eqref{prf:light_2file_upper} completes the proof of\eqref{eq:asymp:compare:equal_2}.
    
    First, we prove \eqref{prf:light_2file_lower} and lower bound the response time of the coded system. For this proof, we use \Cref{lem:1-p1} that states that $\probab_{i0} = 1 - O(\numcoded/n)$, for $i=1, 2$. Then, the arrival rate of tasks to any server in $\set_i$ is bounded as follows 
    
    \begin{equation}\label{eq:2file_lower_nu1}
        \psarrival_i \geq\frac{ \probab_{i0}\arrival_i}{\numserver_i} = \frac{\left(1 - O\left(\frac{\numcoded}{n}\right)\right)(\allocation_i n - \slack_i)}{\allocation_i(n - \numcoded)}.
    \end{equation} 
    Using these, we lower bound the response time of the coded system as follows
    \begin{align}
        \responsetime_{\text{coded}} &=  \sum_{i = 1}^k\sum_{j = 0}^{k}\frac{\arrival_i \probab_{ij}}{\sum_{\ell = 1}^k \arrival_\ell} \times \left(\text{Mean response time of jobs served using $j$ coded servers}\right)\\
        &\geq  \sum_{i = 1}^k\frac{\arrival_i \probab_{i0}}{\arrival_1 + \arrival_2} \times \left(\text{Mean response time of jobs served using $0$ coded servers}\right)\\
        &=  \sum_{i = 1}^k\frac{\arrival_i \probab_{i0}}{\sum_{\ell = 1}^k \arrival_\ell}\frac{1}{1 - \psarrival_i}\\
        &\geq  \sum_{i = 1}^k\left(1 - O\left(\frac{\numcoded}{n}\right)\right)\left(\frac{\arrival_i }{\sum_{\ell = 1}^k \arrival_\ell}\frac{1}{1 - \frac{(1 - O(\numcoded/n))(\allocation_i n - \slack_i)}{\allocation_i(n - \numcoded)}}\right)\label{eq:2file_light_lower:1}\\
        &\geq \sum_{i = 1}^k \left(1 - O\left(\frac{\numcoded}{n}\right)\right)\left(\frac{\arrival_i }{\sum_{\ell = 1}^k \arrival_\ell}\frac{1}{1 - (1 - O(\numcoded/n))(1 - (\slack_i/\allocation_in))(1 + (\numcoded/n))}\right)\label{eq:2file_light_lower:2}\\
        &=  \sum_{i = 1}^k\left(1 - O\left(\frac{\numcoded}{n}\right)\right)\left(\frac{\arrival_i }{\sum_{\ell = 1}^k \arrival_\ell}\frac{1}{\frac{\slack_i}{\allocation_i n} + O\left(\frac{\numcoded}{n}\right)}\right)\label{eq:2file_light_lower:3}\\
        &\geq  \sum_{i = 1}^k\left(1 - O\left(\frac{\numcoded}{n}\right)\right)\left(\frac{\arrival_i }{\sum_{\ell = 1}^k \arrival_\ell}\frac{\allocation_i n}{\slack_i}\left(1 +  O\left(\frac{\numcoded}{\slack_i}\right)\right)\right)\label{eq:2file_light_lower:4}\\
        &= \sum_{i = 1}^k \left(1 - O\left(\frac{\numcoded}{n}\right)\right)\left(\frac{\arrival_i }{\sum_{\ell = 1}^k \arrival_\ell}\frac{\allocation_i n}{\slack_i} +  O\left(\frac{n\numcoded}{(\slack_i)^2}\right)\right)\label{eq:2file_light_lower:5}\\
        &=  \sum_{i = 1}^k\left(1 - O\left(\frac{\numcoded}{n}\right)\right)\left(\frac{\arrival_i }{\sum_{\ell = 1}^k \arrival_\ell}\frac{\allocation_i n}{\slack_i} +  o(1)\right)\label{eq:2file_light_lower:6}\\
        &=  \sum_{i = 1}^k\frac{\arrival_i}{\sum_{\ell = 1}^k \arrival_\ell}\frac{\allocation_i n}{\slack_i} + o(1)\label{eq:2file_light_lower:7}\\
        &=  \responsetime_{\text{uncoded}} + o(1),
    \end{align}
    where
    \begin{itemize}
        \item \eqref{eq:2file_light_lower:1} uses \eqref{eq:2file_lower_nu1} and, \Cref{lem:1-p1}.
        \item \eqref{eq:2file_light_lower:2} uses $1 + (\numcoded/n) \leq (1/(1 - (\numcoded/n)) )$ for any $0 \leq \numcoded \leq n$.
        \item \eqref{eq:2file_light_lower:3} uses the fact that for $i=1, 2$
        \begin{equation}
            O\left(\frac{\numcoded}{n}\right) + O\left(\frac{\slack_i\numcoded}{n^2}\right) + O\left(\left(\frac{\numcoded}{n}\right)^2\right) + O\left(\frac{\slack_i\left(\numcoded\right)^2}{n^3}\right) = O\left(\frac{\numcoded}{n}\right),
        \end{equation}
        since $\numcoded = o(n)$ and $\slack_i = O(n)$ for all $i$.
        \item \eqref{eq:2file_light_lower:6} and, \eqref{eq:2file_light_lower:7} uses the fact that $\slack_i\leq  \slack_{i + 1}$ for all $i$ and $\slack_1 = \omega\left(\sqrt{n \numcoded}\right)$.
    \end{itemize}
    To complete the proof, we now prove \eqref{prf:light_2file_upper} and upper bound the response time. To upper bound the response time of the coded system, we show that there exists a routing probability for which $\responsetime_{\text{coded}} \leq  \responsetime_{\text{uncoded}} +  o(1)$. Consider the routing probability $\probab_{i0} = 1$ for all $i$. The routing probability given above is a sub-optimal routing policy as it never uses the coded servers. Hence
    \begin{align}
        \responsetime_{\text{coded}} &=  \sum_{i = 1}^k\sum_{j = 0}^{k}\frac{\arrival_i \probab_{ij}}{\sum_{\ell = 1}^k \arrival_\ell} \times \left(\text{Mean response time of jobs served using $j$ coded servers}\right)\\
        &\leq  \sum_{i = 1}^k\frac{\arrival_i}{\sum_{\ell = 1}^k \arrival_\ell}\frac{1}{1 - \frac{\allocation_i n - \slack_i}{\allocation_1(n - \numcoded)}}\\
        &\leq  \sum_{i = 1}^k\frac{\arrival_i}{\sum_{\ell = 1}^k \arrival_\ell}\frac{1}{1 - \left(1 - \frac{\slack_i}{\allocation_i n}\right)\left(1 + \frac{2\numcoded}{n}\right)}\label{eq:2file_light_higher:1}\\
        &=  \sum_{i = 1}^k\frac{\arrival_i}{\sum_{\ell = 1}^k \arrival_\ell} \frac{1}{\frac{\slack_i}{\allocation_i n} - \frac{2\numcoded}{n} + \frac{2\slack_i\numcoded}{\allocation_i n^2}}\\
        &\leq  \sum_{i = 1}^k\frac{\arrival_i}{\sum_{\ell = 1}^k \arrival_\ell}\frac{\allocation_i n}{\slack_i}\left(1 +  O\left(\frac{\numcoded}{\slack_i}\right)\right)\\
        &=  \sum_{i = 1}^k\frac{\arrival_i}{\sum_{\ell = 1}^k \arrival_\ell}\frac{\allocation_i n}{\slack_i} + \sum_{i = 1}^k O\left(\frac{n\numcoded}{\left(\slack_i\right)^2}\right)\\
        &=  \responsetime_{\text{uncoded}} + o(1),\label{eq:2file_light_higher:2}
    \end{align}
    where 
    \begin{itemize}
        \item \eqref{eq:2file_light_higher:1} uses $(1/(1 - (\numcoded/n)) \leq 1 + (2\numcoded/n)$ for $\numcoded = o(n)$, and large enough $n$.
        \item \eqref{eq:2file_light_higher:2} uses the fact that $\slack_i\leq  \slack_{i + 1}$ for all $i$ and $\slack_1 = \omega\left(\sqrt{n \numcoded}\right)$.
    \end{itemize}
    This completes the proof for the light region. 
\end{proof}
\subsection{The Inner-heavy Regime}
\begin{proof}[Proof of Theorem~\ref{thm:rt_analyze_general} in the Inner-heavy Regime]
    The inner-heavy regime is the set of arrival rate vectors that satisfy $\slack_i \leq \slack_j$ for any $i\leq j$, $\slack_1 = \Omega(\numcoded)$, $\slack_{1} = o\left(\sqrt{n \numcoded}\right)$, and $\slack_{\bottleneck + 1} = \omega\left(\slack_1\right)$. For this regime, we prove that 
    \begin{equation}
        \responsetime_{\text{coded}} \leq \responsetime_{\text{uncoded}} - \omega(1).
    \end{equation}
    
    The key idea to upper bound the response time is to use a \emph{reasonable} routing policy that distributes the load uniformly across all servers. From our assumption $\arrival_i = \Theta(n)$, we know that arrival rate of tasks to the systematic servers is $\Theta(1)$. Hence, we want the arrival rate of tasks to the coded servers to be $\Theta(1)$. From \Cref{lem:1-p1}, we know that $1 - \probab_{i0} = O(\numcoded/n)$ for all $i$. However, to ensure that the arrival rate of tasks to the coded server is $\Theta(1)$, we want $1 - \probab_{i0} = \Theta(\numcoded/n)$ for at least some $i$. 
    Let $\bottlenecktwo$ be the $\max k$ s.t. $k\leq \bottleneck$ and $\slack_k = O\left(\sqrt{n\numcoded}\right)$.
    
    Since $\slack_i \leq \slack_j$ for all $i\leq j$, the systematic servers of type $i$, for $i \in \{1, \dots, \bottlenecktwo\}$, experience heavy traffic. Hence, we want to decrease their load by serving some of those job types using the coded servers. Therefore, we would like to choose the routing probability where $1 - \probab_{i0} = \Theta(\numcoded/n)$, for all $i\leq \bottlenecktwo$. Consider the following routing probability. Define, 
    \begin{equation}\label{eq:v}
        v = \frac{1}{\bottlenecktwo \sum_{j = 1}^{\bottlenecktwo} \allocation_{j}}
    \end{equation}
    For all jobs of type $i$, such that $i > \bottlenecktwo$,
    \begin{equation}\label{eq:routing_probab_heavy_2file1}
        \probab_{ij} = \begin{cases} 
          1 & j =  0 \\
          0 & \text{else} 
      \end{cases}
    \end{equation}
    For all jobs of type $i$, such that $i \leq \bottlenecktwo$,
    \begin{equation}\label{eq:routing_probab_heavy_2file2}
        \probab_{ij} = \begin{cases} 
          1 - \frac{v\numcoded}{n} & j = 0 \\
          \frac{v\numcoded}{n} & j =  \bottleneck \\
          0 & \text{else} 
      \end{cases}
    \end{equation}
    Moreover, when serving a job of type $i$, for $i \leq \bottlenecktwo$, according to routing probability $\probab_{1\bottleneck}$, only systematic servers in $\set_j$, $j > \bottleneck$ are used. Then, $\psarrival_i$ and $\psarrival_{\text{coded}}$ the arrival rate to servers in  $\set_i$, for all $i \leq k$ and, $\set_{\text{coded}}$ servers respectively, are bounded as follows. For all $i \leq \bottlenecktwo$
    \begin{align}
        \psarrival_i &= \frac{\left(1 - \frac{ v\numcoded}{n}\right)\left(\allocation_i n - \slack_i\right)}{\allocation_i(n - \numcoded)} \\
        &= \frac{1 - \frac{\slack_i}{\allocation_i n} - \frac{ v\numcoded}{ n} + o\left(\frac{ \numcoded}{n}\right)}{1 - \frac{\numcoded}{n}}\\
        &\leq \left(1 - \frac{\slack_i}{\allocation_i n} - \frac{ v\numcoded}{ n} + o\left(\frac{ \numcoded}{n}\right)\right)\left(1 + \frac{1 + v}{2}\frac{\numcoded}{ n}\right)\label{eq:bound_nu_2file_set1}\\
        &= 1 - \frac{\slack_i}{\allocation_i n} - \frac{ v - 1}{2}\frac{\numcoded }{n} + o\left(\frac{\numcoded}{n}\right)\label{eq:bound_nu_2file_set1_final},
    \end{align}
    In the uncoded system, the servers in $\set_i$ for $i\leq \bottlenecktwo$ experiences a higher traffic and arrival rate of tasks in those servers is $(1 - O(\slack_i/n))$. However, in the coded system, because we distribute the load more uniformly, we decrease the arrival rate to $(1 - O(\slack_i/n) - O(\numcoded/n))$. For all $i$ s.t. $\bottlenecktwo < i \leq \bottleneck$, the arrival rate is bounded as 
    \begin{align}
        \psarrival_i &= \frac{\allocation_i n - \slack_i}{\allocation_i(n - \numcoded)} \\
        &= \frac{1 - \frac{\slack_i}{\allocation_i n}}{1 - \frac{\numcoded}{n}}\\
        &\leq \left(1 - \frac{\slack_i}{\allocation_i n}\right)\left(1 + \frac{1 + v}{2}\frac{\numcoded}{ n}\right)\label{eq:bound_nu_2file_set1_mid}\\
        &= 1 - \frac{\slack_i}{\allocation_i n} + O\left(\frac{\numcoded}{n}\right)\label{eq:bound_nu_2file_setmid_final},
    \end{align}
    where \eqref{eq:bound_nu_2file_set1} uses $1/(1 - x) \leq (1 + (1 + \epsilon)x)$, for any $x, \epsilon >0$ when $x = o(1)$. The bound for these servers is similar to the light arrival regime, i.e., although the arrival rate is slightly higher compared to the uncoded system, the effect on the mean response time is $o(1)$. Similarly, for all $i > \bottleneck$
    \begin{align}
        \psarrival_i &= \frac{\allocation_i n - \slack_i + \sum_{j = 1}^{\bottlenecktwo}\frac{v\numcoded}{ n}\left(\allocation_j n - \slack_j \right)}{\allocation_i(n - \numcoded)} \\
        &= \frac{1 - \frac{\slack_i}{\allocation_i n} + \frac{v\numcoded}{\allocation_i n}\sum_{j = 1}^{\bottlenecktwo}\allocation_j - o\left(\frac{ \numcoded}{n}\right)}{1 - \frac{\numcoded}{n}} \\
        &\leq \left(1 - \frac{\slack_i}{\allocation_i n} + \frac{v\numcoded}{\allocation_i n}\sum_{j = 1}^{\bottlenecktwo}\allocation_j - o\left(\frac{ \numcoded}{n}\right)\right)\left(1 + \frac{2\numcoded}{n}\right) \\
        &= 1 - \frac{\slack_i}{\allocation_i n} + \Theta\left(\frac{\numcoded}{n}\right)\label{eq:bound_nu_2file_set2_final},
    \end{align}
    Because of uniform load balancing, the arrival rate of tasks to the servers in $\set_i$ for all $i > \bottleneck$ increases, but the order of the load still remains the same. Although the traffic loss for servers of heavy job types is similar to the traffic gain for servers of light jobs, their effect on the response time is not the same. Finally, for the coded servers,
    \begin{align}
        \psarrival_{\text{coded}} &= \frac{\bottlenecktwo\sum_{j = 1}^{\bottlenecktwo}\frac{v \numcoded}{n}\left(\allocation_j n - \slack_j\right)}{\numcoded}\\
        &= \sum_{j = 1}^{\bottlenecktwo}\left(v \bottlenecktwo \allocation_j  - \frac{\bottlenecktwo v \slack_j}{n}\right)\\
        &= 1 - \sum_{j = 1}^{\bottlenecktwo}\frac{\bottlenecktwo v \slack_j}{n}\\
        &\leq 1 - \frac{(\bottlenecktwo )^2v \slack_1}{n}\label{eq:bound_nu_2file_setC_final}.
    \end{align}
    The coded servers experience higher traffic because of our choice of routing policy. However, because we use the coded servers with $O(\numcoded/n) = o(1)$ probability, the impact on the mean response time is small. Using this bound on the arrival rate of tasks, we bound the mean response time of the coded system as follows. 
    \begingroup
    \allowdisplaybreaks
    \begin{align}
        \responsetime_{\text{coded}} &= \sum_{i = 1}^k\sum_{j = 0}^{k}\frac{\arrival_i \probab_{ij}}{\arrival_1 + \arrival_2} \times \left(\text{Mean response time of jobs served using $j$ coded servers}\right)\\
        &\leq  \sum_{i=1}^{ \bottlenecktwo}\frac{\arrival_i}{\sum_{\ell = 1}^k \arrival_\ell}\left(\frac{1 - \frac{v\numcoded}{n}}{1 - \nu_i^\suppn} +  \frac{v\numcoded}{n}\left(\sum_{\ell = \bottleneck + 1}^k\frac{1}{1 - \nu_\ell^\suppn} + \frac{\bottleneck}{1 - \psarrival_{\text{coded}}}\right)\right) + \nonumber\\
        &\qquad +  \sum_{i=\bottlenecktwo + 1}^{ \bottleneck}\frac{\arrival_i}{\sum_{\ell = 1}^k \arrival_\ell}\frac{1}{1 - \nu_i^\suppn} + \sum_{i=\bottleneck + 1}^{ k}\frac{\arrival_i}{\sum_{\ell = 1}^k \arrival_\ell}\frac{1}{1 - \nu_i^\suppn}\label{eq:bound_rt_2file_heavy:1}\\
        &\leq  \sum_{i=1}^{ \bottleneck}\frac{\arrival_i}{\sum_{\ell = 1}^k \arrival_\ell}\left(\frac{1}{\frac{\slack_i}{\allocation_i n} + \frac{ v - 1}{2}\frac{\numcoded }{n} + o\left(\frac{\numcoded}{n}\right)} +  \frac{v\numcoded}{n}\left(\sum_{\ell = \bottleneck + 1}^k\frac{1}{\Theta\left(\frac{\slack_\ell}{ n}\right)} + \frac{\bottleneck}{\frac{(\bottleneck)^2v \slack_1}{n}}\right)\right) + \nonumber\\
        &\qquad + \sum_{i = \bottlenecktwo + 1}^{\bottleneck}\frac{\arrival_i}{\sum_{\ell = 1}^k \arrival_\ell} \frac{1}{\frac{\slack_i}{\allocation_i n} - O\left(\frac{\numcoded}{n}\right)} + \sum_{i=\bottleneck + 1}^{ k}\frac{\arrival_i}{\sum_{\ell = 1}^k \arrival_\ell}\frac{1}{\frac{\slack_i}{\allocation_i n} - \Theta\left(\frac{\numcoded}{n}\right)}\\
        &\leq  \sum_{i=1}^{ \bottleneck}\frac{\arrival_i}{\sum_{\ell = 1}^k \arrival_\ell}\frac{\allocation_i n}{\slack_i}\left(\frac{1}{1 + \frac{ \allocation_i(v - 1)}{2}\frac{\numcoded }{\slack_i} + o\left(\frac{\numcoded}{\slack_i}\right)}\right) + \Theta\left(\frac{\numcoded}{\slack_{1}}\right) + \nonumber\\
        &\qquad + \sum_{i = \bottlenecktwo + 1}^{\bottleneck}\frac{\arrival_i}{\sum_{\ell = 1}^k \arrival_\ell}\frac{\allocation_i n}{\slack_i}\frac{1}{1 -  O\left(\frac{\numcoded}{\slack_i}\right)}  + \sum_{i=\bottleneck + 1}^{ k}\frac{\arrival_i}{\sum_{\ell = 1}^k \arrival_\ell}\frac{\allocation_i n}{\slack_i}\left(\frac{1}{1 - \Theta\left(\frac{\numcoded}{\slack_i}\right)}\right)\\
        &\leq  \sum_{i=1}^{ \bottleneck}\frac{\arrival_i}{\sum_{\ell = 1}^k \arrival_\ell}\frac{\allocation_i n}{\slack_i}\left(1 - \frac{ \allocation_i(v - 1)}{4}\frac{\numcoded }{\slack_i} - o\left(\frac{\numcoded}{\slack_i}\right)\right) + \Theta\left(\frac{\numcoded}{\slack_{1}}\right) + \nonumber\\
        &\qquad + \sum_{i = \bottlenecktwo + 1}^{\bottleneck}\frac{\arrival_i}{\sum_{\ell = 1}^k \arrival_\ell}\frac{\allocation_i n}{\slack_i}\left(1 +  O\left(\frac{\numcoded}{\slack_i}\right)\right) + \sum_{i=\bottleneck + 1}^{ k}\frac{\arrival_i}{\sum_{\ell = 1}^k \arrival_\ell}\frac{\allocation_i n}{\slack_i}\left(1 + \Theta\left(\frac{\numcoded}{\slack_i}\right)\right)\label{eq:bound_rt_2file_heavy:2}\\
        &\leq  \sum_{i=1}^{ k}\frac{\arrival_i}{\sum_{\ell = 1}^k \arrival_\ell}\frac{\allocation_i n}{\slack_i} - \sum_{i = 1}^{\bottlenecktwo}\Theta\left(\frac{n\numcoded}{\left(\slack_i\right)^2}\right) + \sum_{i = \bottlenecktwo + 1}^{k}\Theta\left(\frac{n\numcoded}{\left(\slack_{i}\right)^2}\right)\\
        &\leq \responsetime_{\text{uncoded}} - \omega(1),\label{eq:bound_rt_2file_heavy:3},
    \end{align}
    where
    \begin{itemize}
        \item \eqref{eq:bound_rt_2file_heavy:1} upper bounds mean of the max response time of tasks by the mean of their sum,
        \item \eqref{eq:bound_rt_2file_heavy:2} uses $1/(1 - x) \leq (1 + (1 + \epsilon)x)$, for any $x=o(1)$ and any $\epsilon>0$,
        \item \eqref{eq:bound_rt_2file_heavy:3} uses $\slack_1 = o\left(\sqrt{n\numcoded}\right)$. Moreover, from the definition of $\bottlenecktwo$, if $\bottlenecktwo < \bottleneck$, then $\slack_j = \omega\left(\sqrt{n\numcoded}\right)$ for all $j > \bottlenecktwo$ in which case the statement is true. If $\bottlenecktwo = \bottleneck$, then we use the fact that $\slack_{1} = o(\slack_{\bottleneck + 1})$.
    \end{itemize}
    \endgroup
\end{proof}
\subsection{The Outer-heavy Regime}
\begin{proof}[Proof of Theorem~\ref{thm:rt_analyze_general} in the Outer-heavy Regime]
    The outer-heavy regime is the set of arrival rate vectors $\arrival_i$ that satisfies the property that $\slack_i \leq \slack_j$ for all $i\leq j$ and $\slack_i = o(\numcoded)$ and, $\slack_{\bottleneck + 1} = \omega(\numcoded)$ . For this regime, we prove that 
    \begin{equation}
        \responsetime_{\text{coded}} = o\left(\responsetime_{\text{uncoded}}\right).
    \end{equation}
    The proof for the outer-heavy regime is similar to the proof of inner-heavy regime, i.e., we figure out a good routing probability that distributes the load uniformly across the various servers. We consider a similar routing probability that we considered for the inner-heavy regime. We define $\bottlenecktwo$ be the $\max k$ s.t. $k\leq \bottleneck$ and $\slack_k = O(\numcoded)$. Define, 
    \begin{equation}
        v = \frac{1}{\bottlenecktwo \sum_{j = 1}^{\bottlenecktwo} c_{j}}
    \end{equation}
    For all jobs of type $i$ such that $i > \bottlenecktwo$,
    \begin{equation}
        \probab_{ij} = \begin{cases} 
          1 & j =  0 \\
          0 & \text{else} 
      \end{cases}
    \end{equation}
    For all jobs of type $i$ such that $i \leq \bottlenecktwo$,
    \begin{equation}
        \probab_{ij} = \begin{cases} 
          1 - \frac{v\numcoded}{n} & j = 0 \\
          \frac{v\numcoded}{n} & j =  \bottleneck \\
          0 & \text{else} 
      \end{cases}
    \end{equation}
    For the routing probability given above, we can reuse the bound for $\psarrival_i$ and $\psarrival_{\text{coded}}$ given in \eqref{eq:bound_nu_2file_set1_final}, \eqref{eq:bound_nu_2file_setmid_final}, \eqref{eq:bound_nu_2file_set2_final} and, \eqref{eq:bound_nu_2file_setC_final} respectively, i.e., for all $i$ s.t. $i \leq \bottlenecktwo$, 
    \begin{equation}
        \psarrival_i \leq 1 - \frac{\slack_i}{\allocation_i n} - \frac{ v - 1}{2}\frac{\numcoded }{n} + o\left(\frac{\numcoded}{n}\right).
    \end{equation}
    For all $i$ s.t. $\bottlenecktwo < i \leq \bottleneck$, 
    \begin{equation}
        \psarrival_i \leq  1 - \frac{\slack_i}{\allocation_i n} + O\left(\frac{\numcoded}{n}\right).
    \end{equation}
    For all $i$ s.t. $\bottleneck < i$, 
    \begin{equation}
        \psarrival_i \leq 1 - \frac{\slack_i}{\allocation_i n} + \Theta\left(\frac{\numcoded}{n}\right),
    \end{equation}
    For the coded servers, 
  \begin{equation}
        \psarrival_{\text{coded}} \leq 1 - \frac{(\bottlenecktwo )^2v \slack_1}{n}.
    \end{equation}
    \begin{equation}
        \psarrival_{\text{coded}} = 1 - \frac{\slack_1}{\allocation_1 n}.
    \end{equation}
    The above bounds only use the fact that $\numcoded = o(n)$ and $\slack_i = O(\numcoded)$ for all $i\leq \bottlenecktwo$, which is also true for the outer-heavy regime. However, one key difference to note is that unlike the inner-heavy regime, for the outer-heavy regime there is an order-wise decrease in traffic experienced by the systematic servers of type $i \in \{1, 2,\dots, \bottlenecktwo\}$. The traffic decreased from $(1 - O(\slack_i/n))$ to $(1 - O(\numcoded/n))$. This enables the coded system to achieve an order-wise improvement in the mean response time. The response time of the coded system is then upper bounded as follows.
    \begingroup
    \allowdisplaybreaks
    \begin{align}
        \responsetime_{\text{coded}} &= \sum_{i = 1}^k\sum_{j = 0}^{k}\frac{\arrival_i \probab_{ij}}{\arrival_1 + \arrival_2} \times \left(\text{Mean response time of jobs served using $j$ coded servers}\right)\\
        &\leq  \sum_{i=1}^{ \bottlenecktwo}\frac{\arrival_i}{\sum_{\ell = 1}^k \arrival_\ell}\left(\frac{1 - \frac{v\numcoded}{n}}{1 - \nu_i^\suppn} +  \frac{v\numcoded}{n}\left(\sum_{\ell = \bottleneck + 1}^k\frac{1}{1 - \nu_\ell^\suppn} + \frac{\bottleneck}{1 - \psarrival_{\text{coded}}}\right)\right) + \nonumber\\
        &\qquad +  \sum_{i=\bottlenecktwo + 1}^{ \bottleneck}\frac{\arrival_i}{\sum_{\ell = 1}^k \arrival_\ell}\frac{1}{1 - \nu_i^\suppn} + \sum_{i=\bottleneck + 1}^{ k}\frac{\arrival_i}{\sum_{\ell = 1}^k \arrival_\ell}\frac{1}{1 - \nu_i^\suppn}\label{eq:bound_rt_2file_midheavy:1}\\
        &\leq  \sum_{i=1}^{ \bottlenecktwo}\frac{\arrival_i}{\sum_{\ell = 1}^k \arrival_\ell}\left(\frac{1}{\frac{\slack_i}{\allocation_i n} + \frac{ v - 1}{2}\frac{\numcoded }{n} + o\left(\frac{\numcoded}{n}\right)} +  \frac{v\numcoded}{n}\left(\sum_{\ell = \bottleneck + 1}^k\frac{1}{\Theta\left(\frac{\slack_\ell}{ n}\right)} + \frac{\bottleneck}{\frac{(\bottleneck)^2v \slack_1}{n}}\right)\right) + \nonumber\\
        &\qquad + \sum_{i = \bottlenecktwo + 1}^{\bottleneck}\frac{\arrival_i}{\sum_{\ell = 1}^k \arrival_\ell} \frac{1}{\frac{\slack_i}{\allocation_i n} - O\left(\frac{\numcoded}{n}\right)} + \sum_{i=\bottleneck + 1}^{ k}\frac{\arrival_i}{\sum_{\ell = 1}^k \arrival_\ell}\frac{1}{\frac{\slack_i}{\allocation_i n} - \Theta\left(\frac{\numcoded}{n}\right)}\\
        &\leq  \sum_{i=1}^{ \bottlenecktwo}\frac{\arrival_i}{\sum_{\ell = 1}^k \arrival_\ell}\Theta\left(\frac{n}{\numcoded}\right) + \Theta\left(\frac{\numcoded}{\slack_{1}}\right) + \nonumber\\
        &\qquad + \sum_{i = \bottlenecktwo + 1}^{\bottleneck}\frac{\arrival_i}{\sum_{\ell = 1}^k \arrival_\ell}\frac{\allocation_i n}{\slack_i}\frac{1}{1 -  o(1)}  + \sum_{i=\bottleneck + 1}^{ k}\frac{\arrival_i}{\sum_{\ell = 1}^k \arrival_\ell}\frac{\allocation_i n}{\slack_i}\left(\frac{1}{1 - o(1)}\right)\\
        &\leq  \sum_{i=1}^{ \bottlenecktwo}\frac{\arrival_i}{\sum_{\ell = 1}^k \arrival_\ell}\Theta\left(\frac{n}{\numcoded}\right) + \sum_{i = \bottlenecktwo + 1}^{k}\frac{\arrival_i}{\sum_{\ell = 1}^k \arrival_\ell}\frac{\allocation_i n}{\slack_i}(1 -  o(1))\\
        &\leq  \sum_{i=1}^{ \bottlenecktwo}o\left(\frac{n}{\slack_i}\right)\\
        &\leq  o\left(\responsetime_{\text{uncoded}}\right)\label{eq:bound_rt_2file_midheavy:2}
    \end{align}
    where
    \begin{itemize}
        \item \eqref{eq:bound_rt_2file_midheavy:1} upper bounds the maximum of response times of tasks by their sum.
        \item \eqref{eq:bound_rt_2file_midheavy:2} uses the fact that $\responsetime_{\text{coded}} = O(n/\slack_1)$. This is because of $\slack_1 = o(\numcoded)$ and \eqref{eq:mrt_2_un}.
    \end{itemize}
    Hence, by including coded servers, the order of the response time decreases from $\Theta(n/\slack_1)$ to $\Theta(n/\numcoded)$.
    \endgroup
\end{proof}
\section{Proof of Sufficient Conditions for Unstable regimes}\label{prf:sufficient}
\subsection{Proof of Sufficient Condition for Uncoded-Unstable Regime}
\begin{proof}
    The region is defined as $\left|\slack_\bottleneck\right| = o(\numcoded)$, $\slack_{\bottleneck + 1} = \omega(\numcoded)$ and $\slack_\bottleneck < 0, \slack_{\bottleneck + 1} \geq 0$. Clearly, the uncoded system is unstable in this regime. To prove the theorem, it suffices to show that the coded system is stable. Let $\bottlenecktwo = \max j$ such that $j \leq \bottleneck$ and $\slack_j < 0$. Then, all jobs of type $\bottleneck + 1, \bottleneck + 2, \dots, k$ can be served by their systematic servers. For a job of type $i$ for $i \leq \bottlenecktwo$, we send $\allocation_i(n - \numcoded)$ amount of traffic to their systematic servers. For the remaining traffic, we send $\bottleneck$ tasks to the coded server and $(k - \bottleneck)$ tasks to servers, one each to $\set_{\bottleneck + 1}, \dots, \set_k$. The servers in set $\set_j$ for all $j\leq \bottlenecktwo$ are stable. 
    
    For $i > \bottleneck$, the total traffic to systematic servers of type $i$ is $\left(\allocation_i n - \slack_i + \sum_{j = 1}^\bottlenecktwo(\allocation_j\numcoded - \slack_j)\right)=\left(\allocation_i n  - \omega\left(\numcoded\right)\right)$ which is less than its total service capacity $\allocation_i(n - \numcoded)$. The traffic to the coded server is $\bottleneck(\sum_{j = 1}^\bottlenecktwo(\allocation_j\numcoded - \slack_j)$ which is less than $\numcoded$ using definition of $\bottleneck$ and $|\slack_1| = o(\numcoded)$. Hence, the system is stable. 
\end{proof}
\subsection{Proof of Sufficient Condition for Coded-Unstable Regime}
\begin{proof}
    Given
    \begin{equation}
        \slack_{\bottleneck + 1} = \omega(\numcoded).
    \end{equation}
    Total service capacity required by files $\slack_1, \dots, \slack_{\bottleneck + 1}$ is $\sum_{i = 1}^{\bottleneck + 1}(\allocation_in - \slack_i)$. Let $p$ be the fraction of the jobs that are serviced completely by servers not dedicated to the first $(\bottleneck + 1)$ servers. For stability, we require
    \begin{equation}
        (1 - p)\sum_{i = 1}^{\bottleneck + 1}(\allocation_in - \slack_i) \leq \sum_{i = 1}^{\bottleneck + 1}\allocation_i(n - \numcoded).
    \end{equation}
    Simplifying the equation , we get, 
    \begin{equation}
        p \geq \frac{\sum_{i = 1}^{\bottleneck + 1}(\allocation_i\numcoded - \slack_i)}{\sum_{i = 1}^{\bottleneck + 1}(\allocation_in - \slack_i)}
    \end{equation}
    Hence, the total service required from the coded servers for the first $(\bottleneck + 1)$ job types is at least 
    \begin{equation}\label{eq:unstable}
        (\bottleneck + 1)\left(\sum_{i = 1}^{\bottleneck + 1}(\allocation_i\numcoded - \slack_i)\right) \geq \numcoded.
    \end{equation}
    where \eqref{eq:unstable} uses the definition of $\bottleneck$ given in \eqref{eq:bottleneck}. Hence, the system is unstable.
\end{proof}

\section{Simulation Setups}\label{sec:app_simulation_setup}
\subsection{Simulations for Fixed Arrival Rate}
\paragraph{{\textbf{Simulation for system with two job types}}}
We consider $n = 2^m$ servers, where we vary $m\in\{6, 7, \cdots, 11\}$. For the uncoded system, we calculate the response time theoretically. We consider a symmetrical allocation of servers, i.e., $\allocation_1 = \allocation_2 = 0.5$. For each experiment, we run the experiment until $10^8$ jobs leave the system and average it over 50 runs to calculate the mean response time. For the coded system, we consider two different values for the number of coded servers $\numcoded = n/16$, and $ \numcoded = \sqrt{n}$. The arrival rate vectors for the various traffic regimes are chosen as follows.
\begin{enumerate}
    \item Light regime: $\bm{\lambda}^\suppn = [n/2 - 5n/32, n/2 - 6n/32]$.
    \item Inner-heavy regime: $\bm{\lambda}^\suppn = [n/2 - (n/2)^{0.55}, n/2 - 6n/32]$.
    \item Outer-heavy regime: $\bm{\lambda}^\suppn = [n/2 - (n/2)^{0.3}, n/2 - 6n/32]$.
\end{enumerate}
\paragraph{{\textbf{Simulation for system with three job types}}}
We consider $n = 3^m$ servers, where we vary $m\in\{5, 7, \cdots, 8\}$. For the uncoded system, we calculate the response time theoretically. We consider an asymmetrical allocation of servers, i.e., $\allocation_1 = 2/11,  \allocation_2 = 0.3/11$, and $\allocation_3 = 6/11$. For each experiment, we run the experiment until $10^8$ jobs leave the system and average it over 50 runs to calculate the mean response time. For the coded system, we consider two different values for the number of coded servers $\numcoded = n/27$, and $ \numcoded = \sqrt{n}$. The arrival rate vectors for the various traffic regimes are chosen as follows.
\begin{enumerate}
    \item Light regime: $\bm{\lambda}^\suppn = [\allocation_1 n - 5n/48, \allocation_2 n - 7n/48, \allocation_3 n - 9n/48]$.
    \item Inner-heavy regime: $\bm{\lambda}^\suppn = [\allocation_1 n - (\allocation_1 n)^{0.55}, \allocation_2 n - (\allocation_2 n)^{0.65}, \allocation_3 n - 9n/48]$.
    \item Outer-heavy regime: $\bm{\lambda}^\suppn = [\allocation_1 n - (\allocation_1 n)^{0.3}, \allocation_2 n - (\allocation_2 n)^{0.7}, \allocation_3 n - 9n/48]$.
\end{enumerate}
\subsection{Simulations for Variable Arrival Rate}
We consider a system with $60$ servers and two job types. For the uncoded system, the $26$ servers are provided to job type $1$ and the remaining to job type $2$. We allocate $7$ coded servers for the coded system and distribute $22$ and $31$ systematic servers to job types $1$ and $2$. For the arrival rate of job type $1$, we consider a pulse wave with a minimum amplitude of $6$ jobs/unit, a maximum amplitude of $18$ jobs/unit, and a period of $2000$ units. Similarly, for the arrival rate of job type $2$, we consider a pulse wave with a minimum amplitude of $12$ and a maximum amplitude of $30$ and a period of $2000$ units. For job type $1$, we consider the same lengths for high amplitude and low amplitude duration. For job type $2$, we consider $1200$ and $800$ units of length for high amplitude and low amplitude duration, respectively. Moreover, we consider a $100$ units of rightward shift in the period of job type $2$ compared to the period of job type $1$.
}{}
\end{document}